\documentclass[%
 reprint,
superscriptaddress,
nofootinbib,
 amsmath,amssymb,
 aps,
]{revtex4-1}

\usepackage{color}
\usepackage{amsmath}
\usepackage{amssymb}
\usepackage{aas_macros}
\usepackage{graphicx}
\usepackage{dcolumn}
\usepackage{bm}
\usepackage{hyperref}
\usepackage{cleveref}
\usepackage{pbox}
\usepackage[utf8]{inputenc}
\usepackage[]{times}

\definecolor{orange}{rgb}{1.0, 0.5, 0.0}

\graphicspath{{./}}

\begin{document}

\interfootnotelinepenalty=10000
\preprint{APS/123-QED}

\title{Dissipative Magnetohydrodynamics for Non-Resistive Relativistic Plasmas:\\
{\it An implicit second-order flux-conservative formulation with stiff relaxation}}

\author{Elias R. Most}
\email{emost@princeton.edu}
\affiliation{Princeton Center for Theoretical Science, Princeton University, Princeton, NJ 08544, USA}
\affiliation{Princeton Gravity Initiative, Princeton University, Princeton, NJ 08544, USA}
\affiliation{School of Natural Sciences, Institute for Advanced Study, Princeton, NJ 08540, USA}

\author{Jorge Noronha}
\email{jn0508@illinois.edu}
\affiliation{Illinois Center for Advanced Studies of the Universe, Department of Physics,
University of Illinois at Urbana-Champaign, Urbana, IL 61801, USA}

\date{\today}

\begin{abstract}
  Based on a 14-moment closure for non-resistive (general-) relativistic viscous
  plasmas, we describe a new numerical scheme that is able to handle all
  first-order dissipative effects (heat conduction, bulk and shear
  viscosities), as well the anisotropies induced by the presence of
  magnetic fields. The latter is parameterized in terms of a thermal
  gyrofrequency or, equivalently, a thermal Larmor radius and allows to
correctly capture the thermal Hall effect.  By solving an
  extended Israel-Stewart-like system for the dissipative quantities that
  enforces algebraic constraints via stiff-relaxation, we are able to cast
  all first-order dissipative terms in flux-divergence form. This allows us
  to apply traditional high-resolution shock capturing methods to the
  equations, making the system suitable for the numerical study of highly
  turbulent flows. We present several numerical tests to assess the
  robustness of our numerical scheme in flat spacetime. 
  The 14-moment closure can seamlessly interpolate
  between the highly collisional limit found in neutron star mergers, and
  the highly anisotropic limit of relativistic Braginskii magnetohydrodynamics appropriate for weakly collisional plasmas in black-hole accretion problems.  We
  believe that this new formulation and numerical scheme will be useful for
  a broad class of relativistic magnetized flows.
\end{abstract}

\maketitle


\section{Introduction}

Relativistic fluid dynamics is widely used as an effective description of
long wavelength, long time phenomena in a variety of many-body systems such
as the quark-gluon plasma  \cite{Romatschke:2017ejr}, the early universe
\cite{Weinberg:2008zzc}, and the hot and ultradense matter formed in
neutron star mergers \cite{Baiotti:2016qnr}. In the context of heavy-ion
collisions, experiments at the Relativistic Heavy Ion Collider  and the
Large Hadron  Collider  have  provided  overwhelming  evidence  that  the
quark-gluon plasma formed in highly energetic nuclear collisions behaves as
a strongly-interacting relativistic fluid over distance scales not much
larger than the size of a nucleus \cite{Heinz:2013th}. Through the last
decade (see \cite{Romatschke:2017ejr} for a review), detailed comparisons
between numerical hydrodynamic simulations and experimental data have shown
that  the  matter  formed  in  such  collisions  is  not a perfect fluid,
i.e., viscous effects are needed to describe experimental data. In fact,
the quantitative extraction of the transport coefficients (such as shear
and bulk viscosities) of the quark-gluon plasma from heavy-ion collision
data
\cite{Bernhard:2019bmu,Nijs:2020roc,JETSCAPE:2020shq,JETSCAPE:2020mzn}
provides key guidance toward understanding the novel out-of-equilibrium
properties that emerge in deconfined quark-gluon matter at extreme
temperatures and moderate baryon densities.

Hydrodynamic simulations of the quark-gluon plasma in heavy-ion collisions
solve second-order relativistic viscous hydrodynamic equations of motion
\cite{Israel:1979wp}, which may be obtained from entropy considerations,
relativistic kinetic theory, or from a resummation of the gradient series
\cite{Baier:2007ix}. Current formulations in use
\cite{Baier:2007ix,Denicol:2012cn} share the same spirit of  Israel-Stewart
theory \cite{Israel:1979wp} in which dissipative corrections to the
energy-momentum tensor and the conserved currents behave as new degrees of
freedom that obey additional (relaxation-type) equations of motion, which
must be solved together with the conservation laws to determine the
evolution of the fluid. Alternative formulations, such as anisotropic
hydrodynamics
\cite{Martinez:2010sc,Florkowski:2010cf,Bazow:2013ifa,Alqahtani:2017mhy,Alqahtani:2017jwl},
have also been used in heavy-ion simulations. 

In the linearized regime around equilibrium, it is known that there are
conditions involving the equation of state and the transport coefficients
which, once fulfilled, guarantee stability and causality of second-order
theories \cite{Hiscock:1983zz,Olson:1990rzl}. In the nonlinear regime
probed in heavy-ion collisions, however, the backreaction from the
dissipative corrections becomes important and more stringent conditions to
ensure causality have been derived \cite{Bemfica:2019cop,Bemfica:2020xym}.
See  \cite{Chiu:2021muk,Plumberg:2021bme} for discussions on the relevance
of such nonlinear causality conditions in the description of the evolution
of the quark-gluon plasma. 

Numerical implementations of  viscous,  Israel-Stewart-like equations of
motion can be found both in Eulerian
\cite{Romatschke:2007mq,Schenke:2011bn,Bozek:2011ua,DelZanna:2013eua,Karpenko:2013wva,Shen:2014vra,Habich:2014jna,Romatschke:2015gxa,Bazow:2016yra,Okamoto:2017ukz}
and also in Lagrangian algorithms
\cite{Noronha-Hostler:2013gga,Noronha-Hostler:2014dqa,Noronha-Hostler:2015coa}.
In the former, numerical methods originally written to solve flux
conservative problems are adapted \cite{Jeon:2015dfa} to determine the
evolution of the system, even though the second-order viscous formulations
\cite{Denicol:2012cn,Baier:2007ix} employ equations of motion that are, in
general, not in flux conservative form. In the latter, a Lagrangian method
based on Smoothed Particle Hydrodynamics \cite{Monaghan:1992rr}, extended
to the relativistic regime  \cite{Aguiar:2000hw}, is used in heavy-ion
simulations
\cite{Noronha-Hostler:2013gga,Noronha-Hostler:2014dqa,Noronha-Hostler:2015coa}.          

In the context of neutron star mergers, effective dissipation is crucial to
model the correct equilibrium of accretion disks \cite{1973A&A....24..337S}
and capture potentially out-of-(weak)-equilibrium effects that could affect
the gravitational wave emission \cite{Alford:2017rxf,Most:2021zvc} or late
time evolution of the remnant
\cite{Radice:2017zta,Fujibayashi:2017puw,Shibata:2017jyf,Fujibayashi:2020qda}.
Moreover, turbulent dynamo processes in the merger
\cite{Price:2006fi,Kiuchi:2015sga,Aguilera-Miret:2020dhz,Skoutnev:2021chg}
critically depend on the presence of a resistive and a viscous scale.  In
the case of Rayleigh-Taylor instabilities potentially present at the
surface of the merging neutron stars, the amplification and growth scale
might depend critically on the shear viscosity \cite{Skoutnev:2021chg}.

Several attempts have been made to incorporate viscosity into astrophysical
fluid dynamics simulations of relativistic systems.  In the context of
black hole accretion and neutron star merger remnants, some have considered simple (and acausal
\cite{Hiscock:1985zz}) Navier-Stokes approaches, e.g. Ref.
\cite{Duez:2004nf,Fragile:2018xee}, whereas more recently Israel-Stewart-like
formulations have been considered in Ref. \cite{Shibata:2017jyf,Chabanov:2021dee}.
Apart from second-order formulations, recent progress in causal and
  hyperbolic first-order theories
  \cite{Bemfica:2017wps,Kovtun:2019hdm,Bemfica:2019knx,Hoult:2020eho,Bemfica:2020zjp}
  has also enabled the first simulations of such theories in the context of
  conformal relativistic fluids \cite{Pandya:2021ief}.
  Another conceptually very different approach with symmetric hyperbolic structure originating from the study
  of continuum mechanism was presented by Refs. \cite{Peshkov:2019syf,Romenski:2019qzs}.
In the 
context of mimicking effective shear viscous dissipation associated with
small-scale magnetic turbulence, large-eddy closures have been proposed
\cite{Radice:2017zta,Radice:2020ids}, although additional effort is
required to ensure covariance of such a closure \cite{Duez:2020lgq}. In the
same spirit, an effective Israel-Stewart shear viscous formulation was
proposed in \cite{Shibata:2017jyf} and successfully applied to long-term
studies of post-merger remnants \cite{Fujibayashi:2017puw}.

In this work we take a different approach in that we consider dissipative
evolution equations derived from a moment expansion of the relativistic
Boltzmann equation \cite{Denicol:2012cn}.  Going beyond all works above, we
include all first-order dissipative effects, such as heat conduction, bulk
and shear viscosity, and also include the anisotropies induced by
magnetic fields in the equations of motion of the dissipative quantities in a fully covariant way.
The only effect omitted here is resistivity, which we will consider in a
forthcoming study.  To this end, we incorporate the gyrofrequency as a free
parameter, which allows us to control the degree of anisotropy of heat
conduction and shear stresses relative to the direction of the (comoving)
magnetic field.  Critically, Ref. \cite{Denicol:2018rbw} has shown that in a
systematic expansion in inverse Reynolds and Knudsen numbers, the
associated magnetic field coupling considered in this work is the only possible term that appears
at first and second order in a gradient expansion, if resistive effects of the plasma are neglected
(see \cite{Denicol:2019iyh} for the resistive case). For extreme magnetic
field strengths relative to the plasma energy, which are not considered
here, additional terms might have to be added at second-order, though
\cite{Panda:2020zhr}.

Additionally, one issue affecting all previous numerical (Eulerian)
implementations of Israel-Stewart formulations is the appearance of source
terms with time derivatives. While the presence of strong shocks and fluid
gradients in turbulent astrophysical scenarios calls for high-resolution
shock capturing (HRSC) schemes, these derivatives are not commonly treated
in this way, see e.g. \cite{DelZanna:2013eua}. 
This can easily lead to
numerical instabilities in the flow in more complex simulations, if the dissipative terms develop
gradients themselves, which is associated with regions of numerically under-resolved
physical viscosity. In this work, we address this point in full, by
carefully designing an HRSC scheme, where all first-order dissipation terms
are treated in flux-divergence form \cite{toro2013riemann}. This is facilitated by the use of
stiff relaxation terms that are treated with implicit time integration
schemes \cite{pareschi2005implicit}. The use of such stiff relaxation terms
is inspired by the implementation of the force-free electrodynamics
limit in numerical relativity simulations
\cite{Alic:2012df,Palenzuela:2012my,Most:2020ami}.  Overall, this makes our
scheme suitable for applications in heavy-ion collisions and astrophysical
fluid dynamics alike.\\

This paper is structured as follows. In Sec.\ \ref{sec:method} we describe
the equations of motion we employ and in Sec.\ \ref{sec:flux_conservative} we provide a detailed derivation of how to recast
them in flux-divergence form. In Sec.\ \ref{sec:numerics}, we describe the
numerical method used to solve these equations, and in particular how we handle stiff source terms.
In Sec.\ \ref{sec:tests} we demonstrate the ability of this scheme to
successfully model relevant test problems, and provide a discussion of the
results in Sec.\ \ref{sec:conclusions}.
Throughout this paper, we adopt geometrical units 
where $G=c=k_B=1$ and a mostly plus signature for the Lorentzian spacetime metric $g_{\mu\nu}$.

\section{Non-Resistive Relativistic Dissipative Magnetohydrodynamics}
\label{sec:method}

A relativistic ideal fluid can be described via its rest-mass (``baryon")
density $\rho$, energy density $e$, and four-velocity $u^\mu$ (normalized
such that $u_\mu u^\mu = -1$). The equilibrium pressure $P$ is defined in
terms of the other thermodynamic variables such that, for instance, $P =
P(\rho,e)$. This defines the equation of state, with which one may compute
the temperature $T=T(\rho,e)$ using standard thermodynamic relations.  The
equilibrium hydrodynamics system can be described by means of an
energy-momentum tensor
\begin{align}
  T^{\mu\nu}_{\rm hydro} = e u^\mu u^\nu + P \Delta^{\mu\nu},
\end{align}
where 
\begin{align}
  \Delta_{\mu\nu} =g_{\mu\nu}+ u_\mu u_\nu ,
  \label{eqn:Delta}
\end{align}
is the rank-2 projector orthogonal to the flow velocity, i.e.
$\Delta^{\mu\nu} u_\mu =0$. Conservation of baryon number is defined in
terms of the rest-mass current $N^\mu = \rho u^\mu$ and baryon number conservation
\begin{equation}
    \nabla_\mu N^\mu = 0.
    \label{conservation_current}
\end{equation}
In this work we will also consider the effects of a (co-moving) magnetic
field  $b^\mu$ tightly coupled to the fluid, i.e. in the limit of infinite 
electric conductivity
\cite{Denicol:2018rbw}. We note that $b^\mu u_\mu = 0$ so that, in the local rest frame of
the fluid where $u^\mu = (1,0,0,0)$, the magnetic field 4-vector only has
nonzero spatial components. 

The electromagnetic field is described by the electromagnetic 
field strength tensor and its dual
\begin{align}
  &F^{\mu\nu}:= \sqrt{b^2} b^{\mu\nu}=  - \varepsilon^{\mu\nu\alpha\beta} u_\alpha b_\beta,\\
    &{^\ast}F^{\mu\nu}= u^\mu b^\nu - b^\mu u^\nu.
\end{align}
We stress again that these expressions only hold in the limit infinite
  electric conductivity,
in which the comoving electric field $e^\mu=0$ vanishes. For later convenience, we have also introduced above the shorthand notation
$b^{\mu\nu}$.
The evolution of the magnetic field is then governed by the Maxwell equation $\nabla_\mu{^\ast}F^{\mu\nu}=0$, which may be written as
\begin{align}
    \nabla_\mu \left(b^\mu u^\nu - b^\nu u^\mu\right) =0.
    \label{magnetic_field_eq}
\end{align}
It is useful to introduce the projector
\begin{align}
    \Xi^{\mu\nu} = \Delta^{\mu\nu} - \frac{1}{b^2}b^\mu b^\nu,
\end{align}
which is orthogonal to $u_\mu$ and $b^\mu$. Using this projector, it can be shown that
\begin{align}
    F^{\mu\alpha} \Xi^{\nu}_{\alpha} =  F^{\mu\alpha} \Delta^{\nu}_{\alpha} =F^{\mu\nu},
    \label{eqn:FX}
\end{align}
and 
\begin{align}
    F^{\mu\alpha} F_{\nu\alpha} = b^2 \Xi^{\mu}_{\nu}. 
    \label{eqn:FF}
\end{align}
The electromagnetic fields give rise to an energy-momentum tensor given by
\cite{Duez:2005sf}
\begin{align}
  T^{\mu\nu}_{\rm EM} = \frac{1}{2} b^2 u^\mu u^\nu + \frac{1}{2}b^2
  \Delta^{\mu\nu} - b^\mu b^\nu.
  \label{eqn:TEM}
\end{align}
In total, the generalized energy-momentum tensor for non-resistive ideal magnetohydrodynamics can be written as
\begin{align}
  T^{\mu\nu} = T^{\mu\nu}_{\rm hydro} + T^{\mu\nu}_{\rm EM}.
  \label{eqn:Tmunu}
\end{align}
The equations of
motion of ideal relativistic magnetohydroynamics stem from the conservation of
energy and momentum
\begin{equation}
    \nabla_\mu T^{\mu\nu}=0,
    \label{conservation_EM}
\end{equation}
coupled to Eq. \eqref{magnetic_field_eq}. 

\subsection{Second-order hydrodynamic equations}

In the following we summarize a formalism for describing out-of-equilibrium
dynamics allowing for viscous corrections to the system.  These can be
encoded by the viscous stress tensor $\Pi_{\mu\nu}$, which is added to the
fluid's energy-momentum tensor. In this work, we use the so-called Eckart
hydrodynamic frame \cite{Eckart:1940te} in which the rest-mass current
remains $N^\mu = \rho u^\mu$, all viscous/dissipative effects appear only
in the expression for the energy-momentum tensor, and there are no
out-of-equilibrium corrections to the energy density.

Considering the number of degrees of freedom, we can see that $T^{\mu\nu}$
and $N^\mu$ have a total of 14 (since $T^{\mu\nu}$ is symmetric), but in
equilibrium only five of them are actually independent (e.g.
$\{e,\rho,u^\mu\}$). Accounting for out-of-equilibrium dynamics of the
system then amounts to prescribing evolution equations for the remaining 9
degrees of freedom.  These can be parameterized in terms of the bulk scalar
pressure $\Pi$, the symmetric anisotropic pressure tensor $\pi^{\mu\nu}$,
and the heat flux $q^\mu$, which obey a distinct set of constraints on the
dissipative momenta and stresses, i.e.
\begin{align}
   q_\mu u^\mu &=0, \label{eqn:constraintsQ} \\
  \pi_{\mu\nu} u^\mu &=0, \label{eqn:constraintsPi1} \\
  \pi^{\mu}_\mu  &=0. \label{eqn:constraintsPi2}
\end{align}
Therefore, once these constraints are fulfilled, the set
$\{\pi_{\mu\nu},\Pi,q_\mu\}$ indeed only has 9 independent degrees of
freedom. As we show in this paper, the way these constraints are imposed
heavily influences the numerical methods required to study this system.
Overall, we can group these contributions into the viscous stress tensor
\begin{align}
  \Pi_{\mu\nu} = q_\mu u_\nu + q_\nu u_\mu + \Pi \Delta_{\mu\nu} + \pi_{\mu\nu},
  \label{eqn:Pimunu}
\end{align}
which is then added to the hydrodynamics equations
\begin{align}
T^{\mu\nu} \rightarrow T^{\mu\nu} + \Pi^{\mu\nu}.
\end{align}
The combined stress-energy tensor, including the dissipative contributions, is covariantly conserved
\begin{align}
   & \nabla_\mu \left(T^{\mu\nu} + \Pi^{\mu\nu}\right)=0,
\end{align}
whereas $N^\mu$ remains unchanged with dynamics given by
\eqref{conservation_current}. We point out that we choose the Eckart frame
in this work solely for numerical convenience in the context of
astrophysical applications. Clearly, other definitions of the hydrodynamic
fields, i.e. other hydrodynamic frames \cite{Israel:1979wp}, could have
been used. For instance, one could have removed the heat flux and
introduced a diffusive correction to the baryon current. Alternatively, one
may employ other frames in which both terms are present. For a deeper
discussion about hydrodynamic frames, and how a judicious choice of
hydrodynamic variables can be useful to formulate causal, stable and
strongly-hyperbolic theories of viscous relativistic fluids at first-order
in derivatives, see
\cite{Bemfica:2017wps,Kovtun:2019hdm,Bemfica:2019knx,Hoult:2020eho,Bemfica:2020zjp}.
Second-order Israel-Stewart-like theories in general hydrodynamic frames
have also been investigated in \cite{Tsumura:2009vm}, and more recently in
\cite{Noronha:2021syv,Rocha:2021lze}. 

Having parameterized the viscous degrees of freedom, we need to prescribe a
set of causal evolution equations for them. In this work, we use equations
of motion obtained from a truncation of suitably defined moments of the
relativistic Boltzmann-Vlasov equation for a gas of charged particles
(without dipole moment or spin) dynamically coupled to a magnetic field.
This approach to obtain the second-order hydrodynamic equations employs a
systematic expansion of the Boltzmann equation in powers of the Knudsen and
inverse Reynolds numbers\footnote{The Knudsen numbers measure the ratio
  between microscopic length scales and the macroscopic scales associated
  with gradients of the conserved quantities. In the approach of
  \cite{Denicol:2012cn}, one denotes quantities such as $\Pi/(e+P)$ as
inverse Reynolds numbers, given that they can be seen as relativistic
generalizations of the ratio between viscous and inertial forces.}, and
this method was originally applied in \cite{Denicol:2012cn} in the
investigation of neutral kinetic systems. The equations of motion, and the
transport coefficients obtained within this formulation, are widely used in
current heavy-ion collision simulations of the quark-gluon plasma (usually
done in the absence of eletromagnetic fields). This formalism was extended
to include non-resistive, comoving magnetic field effects using the
Boltzmann-Vlasov equation in \cite{Denicol:2018rbw}. The
magnetohydrodynamical equations of motion were obtained in the so-called
14-moment approximation \cite{Israel:1979wp} and they essentially provide a
generalization of Israel-Stewart fluid dynamics to the case of
non-vanishing magnetic fields. The full resistive case (which includes
effects from comoving electric and magnetic fields) was worked out later in
this approach in Ref.\ \cite{Denicol:2019iyh}.

In our work, the dissipative variables obey the following advection-type
relaxation equations\footnote{We note that the original derivation of the
  non-resistive equations of motion in  \cite{Denicol:2018rbw} used the
  Landau frame \cite{LandauLifshitzFluids}. The Eckart frame version of
  these equations can be found by taking the non-resistive limit of the
  equations derived in \cite{Denicol:2019iyh}.}
\begin{widetext}
\begin{align}
&\tau_\Pi u^\mu \nabla_\mu \Pi + \delta_{\Pi \Pi} \Pi\, \nabla_\mu u^\mu + \Pi = -\zeta  \nabla_\mu u^\mu
    \label{eqn:IS_bulk} \\
&\tau_q \Delta^\alpha_\nu u^\mu \nabla_\mu q^\nu +  \delta_{qq} q^\alpha \nabla_\mu u^\mu + q^\alpha
=
 -\kappa  T u^\mu \nabla_\mu u^\alpha  - \kappa \Delta^{\alpha\mu}
 \nabla_\mu T -  \delta_{qB} F^{\alpha\mu} q_\mu
    \label{eqn:IS_heat} \\
 &   \tau_\pi \Delta^{\alpha\beta}_{\lambda\nu}  u^\mu \nabla_\mu
    \pi^{\lambda\nu} + \delta_{\pi \pi} \pi^{\alpha\beta}\, \nabla_\mu
    u^\mu + \pi^{\alpha \beta} = -\eta
    \Delta^{\alpha\beta}_{\lambda\nu} \left[ \nabla^\nu u^\lambda +
    \nabla^\lambda u^\nu\right] - \delta_{\pi B} F^{\delta\mu} \pi^\gamma_\mu\, \Delta^{\alpha \beta}_{\gamma\delta} ,
    \label{eqn:IS_shear}
\end{align}
\end{widetext}
where we have introduced the symmetric trace-free projector
\begin{align}
  \Delta^{\alpha\beta}_{\lambda\nu} =\frac{1}{2}\left[ \Delta^{\alpha}_\lambda
  \Delta^{\beta}_\nu + \Delta^{\alpha}_\nu \Delta^{\beta}_\lambda\right] -
  \frac{1}{3} \Delta^{\alpha\beta} \Delta_{\lambda\nu}.
  \label{eqn:tildeDelta}
\end{align}
Above, $\eta$ is the shear viscosity, $\zeta$ is the bulk viscosity,
$\kappa$ is the thermal conductivity, $\tau_\Pi$ is the bulk relaxation
time, $\tau_q$ is the relaxation time for heat flux, while $\tau_\pi$ is
the shear viscosity relaxation time. Besides those coefficients, the
equations of motion also contain the additional second-order transport
coefficients $\{\delta_{\Pi\Pi},\delta_{qq}, \delta_{\pi\pi}\}$ and the new
coefficients that determine the coupling between the dissipative quantities
and the electromagnetic field tensor, $\delta_{qB}$ and $\delta_{\pi B}$.
Expressions for all of these coefficients, including $\delta_{qB}$ and
$\delta_{\pi B}$, in terms of the temperature and chemical potential for an
ultrarelativistic gas can be found in
\cite{Denicol:2012cn,Denicol:2019iyh,Denicol:2018rbw}. In this work, we
treat the ratios $\delta_{\Pi \Pi}/\tau_\Pi$,$\zeta/\tau_\Pi$,
$\delta_{qq}/\tau_q$, $\kappa/\tau_q$, $\delta_{\pi\pi}/\tau_\pi$,
$\eta/\tau_\pi$, $\delta_{\pi B}/\tau_\pi$, $\delta_{q B}/\tau_q$ as free
parameters defined in terms of thermodynamic quantities ($e$ and $\rho$).
It is important to remark that in the formulation of Ref.\
\cite{Denicol:2018rbw} the form of the equations of motion remains very
close to that found in standard Israel-Stewart theory, with additional
terms that couple the fluid variables directly to the magnetic field. Also,
we note that in this first work we have not included all the possible
second-order terms that appear in
\cite{Denicol:2012cn,Denicol:2018rbw,Denicol:2019iyh}. A systematic
investigation of their effects will be given in a forthcoming study.

We note that any second-order theory with relaxation equations for the
dissipative currents given by \eqref{eqn:IS_bulk}, \eqref{eqn:IS_heat}, and
\eqref{eqn:IS_shear} has an asymptotic relativistic Navier-Stokes regime.
This can be seen by performing a systematic expansion in gradients,
together with the conservation laws \cite{Baier:2007ix}. 
In the absence of magnetic fields, to first order in gradients the system
reduces to
\begin{align}
   \Pi &= -\zeta \nabla_\mu u^\mu +\mathcal{O}(\partial^2),\\
   q^\alpha &= -\kappa  T u^\mu \nabla_\mu u^\alpha  - \kappa \Delta^{\alpha\mu}
   \nabla_\mu T +\mathcal{O}(\partial^2), \\
   \pi^{\alpha\beta} &= -\eta \Delta^{\alpha\beta}_{\lambda\nu} \left[ \nabla^\nu u^\lambda + \nabla^\lambda u^\nu\right]
   +\mathcal{O}(\partial^2),
\end{align}
which is the standard relativistic generalization of Navier-Stokes theory
\cite{Eckart:1940te}. It is important to understand that this limit can
never be reached exactly, since it reduces the character of the
hydrodynamic evolution equations from hyperbolic to parabolic, rendering
the system acausal \cite{PichonViscous}. Also, we note that the standard
constraints that stem from a linearized analysis of causality and
stability, derived in the absence of a magnetic field in
\cite{Hiscock:1983zz}, automatically hold for the system of equations we
use. For a linearized study of stability and causality in the presence of a
nonzero magnetic field, see \cite{Biswas:2020rps}. 


\section{Flux-conservative formulation}
\label{sec:flux_conservative}

In the following, we will reformulate the equations for the bulk pressure
\eqref{eqn:IS_bulk}, heat flux \eqref{eqn:IS_heat}, and shear-stress tensor
\eqref{eqn:IS_shear} in a flux-conservative form \cite{toro2013riemann,rezzolla2013relativistic} that is suitable for numerical simulations.
In particular, we will suitably extend the system in a way that allows us to recast all first-order gradient terms into flux-divergence form, i.e. $\nabla_\mu \mathcal{F}^\mu$, for a suitable flux function $\mathcal{F}^\mu$.
To this end, to better illustrate our approach, we consider all viscous effects separately.

\subsection{Bulk pressure}


We start out by considering the Israel-Stewart-like equation
\eqref{eqn:IS_bulk} for the bulk
pressure $\Pi$.
After a simple algebraic manipulation it can be shown that
Eq. \eqref{eqn:IS_bulk} is equivalent to
\begin{align}
  \nabla_\mu \left[ \left( \Pi + \frac{\zeta}{\tau_\Pi} \right) u^\mu
  \right] = &- \frac{1}{\tau_\Pi} \Pi \nonumber \\
  &+ \left(1-
  \frac{\delta_{\Pi\Pi}}{{\tau_\Pi}} \right) \Pi
  \theta + u^\mu \nabla_\mu \frac{\zeta}{\tau_\Pi}.
  \label{eqn:Pi_con}
\end{align}
Here we have introduced the shorthand $\theta = \nabla_\mu u^\mu$. We will provide a detailed treatment of  arbitrary viscous relaxation terms
in Sec. \ref{sec:advection}. We note that depending on the application  this equation can be further simplified. 

In the simplest case of $\zeta/\tau_\Pi =\text{ const}$, or
if $\zeta/\tau_\Pi$ is advected, i.e. $u^\mu \nabla_\mu \left(\zeta/\tau_\Pi\right)=0$ and
$\delta_{\Pi\Pi} = \tau_\Pi$, we find
\begin{align}
  \nabla_\mu \left[ \left( \Pi + \frac{\zeta}{\tau_\Pi} \right) u^\mu
  \right] = - \frac{1}{\tau_\Pi} \Pi,
  \label{eqn:Pi_con_simple}
\end{align}
which, remarkably, is free of derivatives on the right-hand side (RHS).

\subsubsection{Limit of vanishing relaxation time}
\label{sec:zero_diss}

We now consider a particular limit of these equations, showing that
in the limit of vanishing relaxation time $\tau_\Pi\to 0$ the system does
not always approach to a perfect fluid solution.
Indeed, we can see from Eq. \eqref{eqn:Pi_con_simple} that for a choice of
transport coefficient $\zeta = \tau_\Pi\, f\left( \rho,T,\dots\dots
\right)$, with $u^\mu \nabla_\mu f = 0$, the  limit $\tau_\Pi \rightarrow 0$ 
corresponds to the introduction of a newly conserved quantity, i.e.
\begin{align}
  \nabla_\mu \left[ \left( \Pi + \frac{\zeta}{\tau_\Pi} \right) u^\mu
  \right] \simeq 0,
\end{align}
where conservation is to be understood to hold approximately up to second-order corrections.
Physically, this limit is reached when bulk viscous damping happens on
viscous scales 
$\ell_{\rm visc} \gg
\ell_{\rm dyn}$ much larger than the dynamical scale $\ell_{\rm dyn}$.  
We can further see that in this limit, $\Pi+\zeta/\tau_\Pi$ satisfies a continuity equations,
akin to the baryon density $\rho$.
In the absence of effective viscous damping, this
implies that $\Pi$ provides an effective correction to the equation of state, which
now depends on the bulk viscous scalar $P \rightarrow P + \Pi$, and in turn
on the velocity via the continuity equation. This is similar to the use of
microphysical equations of state that depend on compositional information,
given by an advected scalar, such as the electron fraction $Y_e$
\cite{Oertel:2016bki}.

\subsection{Heat conduction}


Next we turn to the equation for heat conduction.
It can trivially be shown that
\begin{align}
  \Delta^\alpha_\nu \left[ 
\tau_q u^\mu \nabla_\mu q^\nu +  \delta_{qq}
q^\nu \, \theta\right]
=& 
 \Delta^\alpha_\nu\left[ -\kappa T u^\mu \nabla_\mu u^\nu 
 \right. \nonumber\\
 &\left. - \kappa \nabla^\nu T
 -\delta_{qB} F^{\nu\mu} q_\mu
\right] - q^\alpha,
  \label{eqn:IS_heat_delta}
\end{align}
where we have used that $u_\alpha \nabla_\mu u^\alpha =0$. The
  specific form of this equation further allows 
  us to add terms proportional to $u^\nu$, since these get projected out by
the global $\Delta^\alpha_\nu$ projection. Along those lines, it will help us
to simplify the equations if we add such a terms in the following way
\begin{align}
  \Delta^\alpha_\nu \left[ 
\tau_q u^\mu \nabla_\mu q^\nu +  \delta_{qq}
q^\nu \, \theta\right]
=& 
 \Delta^\alpha_\nu\left[ -\kappa  u^\mu \nabla_\mu \left( T u^\nu \right)
 \right. \nonumber\\
 & - \kappa \nabla^\nu T 
-\delta_{qB} F^{\nu\mu} q_\mu \nonumber\\
&\left.
- \kappa T \theta u^\nu
- \tau_q T u^\mu u^\nu \nabla_\mu\frac{\kappa}{\tau_q}\right] \nonumber\\
&- q^\alpha .
  \label{eqn:IS_heat_delta2}
\end{align}
We can now understand the meaning of the projector $\Delta^\mu_\nu$ in Eq.\
\eqref{eqn:IS_heat_delta} as follows. 
Contracting Eq.\ \eqref{eqn:IS_heat_delta} with $u_\alpha$, we see that the
sole purpose of the projector is to enforce Eq.\ \eqref{eqn:constraintsQ}, such that $q_\mu$ is orthogonal to $u^\mu$.
In practice, the presence of such a projector will lead to a variety of
gradient terms, which {\it explicitly} impose the
constraint \eqref{eqn:IS_heat_delta}. Such terms are typically numerically
difficult to evaluate and will not only increase the error budget of any
numerical solution, but they will also \textit{numerically} lead to only an
approximate enforcement of Eq.\ \eqref{eqn:constraintsQ}.\\
A better approach, which is commonly used in relativistic force-free
electrodynamics \cite{Alic:2012df,Palenzuela:2012my}, is to instead include the constraint \eqref{eqn:IS_heat_delta}
via stiff relaxation \cite{Alic:2012df}. 
Introducing a new relaxation time $\omega_q \ll \tau_q$, we can
equivalently write
\begin{align}
\tau_q u^\mu \nabla_\mu q^\nu +  \delta_{qq}
q^\nu\, \theta
=&
 -\kappa  u^\mu \nabla_\mu \left( T u^\nu \right)   - \kappa \nabla^\nu T
- \kappa T \theta u^\nu
 \nonumber \\
 &- q^\nu -\delta_{qB} F^{\nu\mu} q_\mu  - \tau_q T u^\mu u^\nu \nabla_\mu\frac{\kappa}{\tau_q} \nonumber \\
& - \frac{\tau_q}{\omega_q} \left( q_\mu u^\mu \right) u^\nu.
  \label{eqn:IS_heat_delta2}
\end{align}
{The effect of this term can best be considered in the Navier-Stokes
  limit (in the absence of magnetic fields), where
  \begin{align}
     q^\mu u_\mu \simeq \omega_q \frac{\kappa}{\tau_q} u^\mu \nabla_\mu T\,.
    \label{eqn:qdamp}
  \end{align}
  Hence, the constraint \eqref{eqn:constraintsQ} will indeed be imposed
when $\omega_q \rightarrow 0$. }
Although Eq. \eqref{eqn:IS_heat_delta2} is more desirable from a numerical
point of view than the original Israel-Stewart equation \eqref{eqn:IS_heat}, we will need implicit
numerical schemes in order to properly compute this term. This will be
explained in detail in Sec. \ref{sec:numerics}. Applying the same reordering that leads to a flux-conservative form
\eqref{eqn:Pi_con} for the bulk scalar we obtain the final form
\begin{align}
  \nabla_\mu \left[ q^\nu u^\mu + \frac{\kappa}{\tau_q} T
  \Delta^{\mu\nu} \right] = &
  -\frac{\delta_{qB}}{\tau_q} F^{\nu\mu} q_\mu
-\frac{1}{\tau_q} q^{\nu} \nonumber\\ 
&+\left[ \left( 1- \frac{\delta_{qq}}{\tau_q}\right) \right]\,
\theta q^\nu \nonumber\\
&+ T g^{\mu\nu} \nabla_\mu
\frac{\kappa}{\tau_q}
 - \frac{1}{\omega_q} \left( q_\mu u^\mu
 \right) u^\nu
 .
  \label{eqn:IS_heat_con}
\end{align}

For a class of theories, where $\kappa/\tau_q =\text{const}$ or advected
$\left( u^\mu \nabla_\mu \left[ \kappa/\tau_q \right]=0 \right)$, and $\delta_{qq}
= \tau_q$, we find the simple flux-conservative
stiff relaxation equation

\begin{align}
  \nabla_\mu \left[ q^\nu u^\mu + \frac{\kappa}{\tau_q} T
  \Delta^{\mu\nu} \right] =& -\frac{1}{\tau_q} q^{\nu}  
-\frac{\delta_{qB}}{\tau_q} F^{\nu\mu} q_\mu \nonumber \\
&  - \frac{1}{\omega_q} \left( q_\mu u^\mu
 \right) u^\nu.
  \label{eqn:IS_heat_simple}
\end{align}

Nonetheless, our numerical scheme is able to solve systems with arbitrary
heat conductivities, see Sec. \ref{sec:advection}.

\subsubsection{Comparison with non-relativistic heat conduction}
It is interesting to consider a simple limiting case in order to highlight
the nature of heat conduction within this formulation. Restricting
ourselves to flat spacetime, zero magnetic field,   and assuming a
vanishing 3-velocity (i.e. $u_i\approx 0$), it can be shown (neglecting
dissipative sources other than heat condution) that the equations reduce to 
\begin{align}
    &\partial_t \left(\rho \varepsilon \right) + \partial_i q^i =0\\
    &\partial_t \left(q^i \right) + \frac{\kappa}{\tau_q} \partial_j\left( \eta^{ij} T \right) = - \frac{1}{\tau_q} q^i.
\end{align}
For simplicity, adopting the ideal gas law $p=\rho \varepsilon \left(\Gamma
-1\right) = \rho \frac{k_B}{m_b} T$, where $k_B$ is the Boltzmann constant
, $m_b$ the baryon mass and $\Gamma$ the adiabatic coefficient, we find
\begin{align}
    &\partial_t T + \frac{m_b\left(\Gamma -1\right)}{k_B \rho}\partial_i
    q^i =0.
\end{align}
In this simple example, we have used that $\partial_t \rho=0$ for vanishing
fluid velocity.  In the Navier-Stokes limit where $q^i \rightarrow - \kappa
\partial^i T$, we find that this expression reduces to the standard heat
equation.  However, because $\tau_q >0$ this limit will not be reached and
instead we find that the temperature evolution obeys
\begin{align}
    \partial_t^2 T - \frac{\kappa m_b\left(\Gamma -1\right)}{k_B \rho
    \tau_q} \Delta T - \frac{1}{\tau_q} \partial_t T  = 0.
    \label{eqn:telegraph}
\end{align}
This is a damped wave equation for the temperature $T$ with wave speed
\begin{align}
    c_q^2 = \frac{\kappa m_b \left(\Gamma -1\right)}{k_B \rho \tau_q} < 1,
\end{align}
where causality places a strict bound on $\kappa/\tau_q$.
This is nothing but the Telegrapher's equation, which has already been
extensively studied in the context of hyperbolic heat conduction (see, e.g. \cite{Romatschke:2009kr}).

\subsection{Shear-viscous stresses}

We finally consider the evolution of the shear stress tensor $\pi^{\alpha
\beta}$, with the aim of recasting it into flux-conservative form. Starting from Eq. \eqref{eqn:IS_shear}, we write
\begin{align}
   \Delta^{\alpha \beta}_{\lambda\nu} \left[ \tau_\pi  u^\mu \nabla_\mu
   \pi^{\lambda\nu}  + 2\eta \nabla^{\left(\alpha\right.}
   u^{\left.\beta\right)}  \right] =& - \delta_{\pi \pi}
\pi^{\alpha\beta}\, \theta - \pi^{\alpha\beta}\nonumber\\
&-\delta_{\pi B} F^{\delta\mu} \pi^\gamma_\mu\, \Delta^{\alpha \beta}_{\gamma\delta}\,,
    \label{eqn:IS_shear_delta}
\end{align}
where $2 \nabla_{\left(\alpha\right.} u_{\left.\beta\right)} = \nabla_\alpha u_\beta + \nabla_\beta u_\alpha $.
Similar to the observations made for Eq.\ \eqref{eqn:IS_heat_delta}, we
find that the introduction of the trace-free projector
$\Delta^{\alpha\beta}_{\lambda\nu}$ was done to ensure the validity
of the constraints \eqref{eqn:constraintsPi1} and \eqref{eqn:constraintsPi2}.
In the same spirit of deriving Eq.\ \eqref{eqn:IS_heat_delta2}, we can
replace the projector by the introduction of a stiff relaxation current.
More specifically, we write
\begin{align}
   \tau_\pi  u^\mu \nabla_\mu
    \pi^{\alpha\beta}  + 2\eta \nabla^{\left(\alpha\right.}
   u^{\left.\beta\right)}  = &- \delta_{\pi \pi}
\pi^{\alpha\beta}\, \theta - \pi^{\alpha\beta} \nonumber\\
& -\delta_{\pi B} F^{\delta\mu} \pi^\gamma_\mu\, \Delta^{\alpha
\beta}_{\gamma\delta}\nonumber \\
&   - 2 \frac{\tau_\pi}{\omega_{\pi}}
  \pi^{\lambda\left(\alpha\right.} u^{\left.\beta\right)} u_\lambda
  \nonumber \\
&  - \frac{\tau_\pi}{\omega_{\pi\pi}} \left(
  \pi^{\mu\nu} g_{\mu\nu}
  \right) g^{\alpha\beta},
    \label{eqn:IS_shear_delta2}
\end{align}
where we have introduced the relaxations times
$\omega_\pi,\omega_{\pi\pi} \ll \tau_\pi$.
Reordering the terms in the same way as for Eqs.\ \eqref{eqn:Pi_con} and
\eqref{eqn:IS_heat_con}, we find
\begin{widetext}
\begin{align}
  \nabla_\mu \left[  \pi^{\alpha\beta} u^\mu  + \frac{\eta}{\tau_\pi} \left( g^{\mu\alpha}
  u^\beta + g^{\mu\beta}  u^\alpha\right) \right] = &\left(1 -
  \frac{\delta_{\pi \pi}}{\tau_\pi}\right) \pi^{\alpha\beta}\, \theta -
  \frac{1}{\tau_\pi}\pi^{\alpha\beta} 
    +  \left( g^{\mu\alpha}
  u^\beta + g^{\mu\beta}  u^\alpha\right) \nabla_\mu \frac{\eta}{\tau_\pi} 
  -\frac{\delta_{\pi B}}{\tau_\pi} F^{\delta\mu} \pi^\gamma_\mu\, \Delta^{\alpha \beta}_{\gamma\delta}
  \nonumber \\ &
   - \frac{1}{\omega_{\pi}}
  \left[ \left(\pi^{\alpha\lambda} u_\lambda\right) u^\beta + \left(\pi^{\beta\lambda}
  u_\lambda\right) u^\alpha \right] - \frac{1}{\omega_{\pi\pi}} \left(
  \pi^{\mu\nu} g_{\mu\nu}
  \right) g^{\alpha\beta}.
    \label{eqn:IS_shear_con}
\end{align}
\end{widetext}
We further note that all terms terms proportional to $u^\mu$ are already accounted for in the constraint damping term and will get removed in the $\omega \rightarrow 0$ limit.
Remarkably, and
differently from the heat conduction and bulk viscosity, this leads to the
removal of the $\nabla_\mu \left(\eta/\tau_\pi\right)$ term from the equations.
Hence, no approximation or other special treatment need to be applied to handle non-constant 
shear viscosity.
The final form of the evolution equation then reads
\begin{widetext}
\begin{align}
  \nabla_\mu \left[  \pi^{\alpha\beta} u^\mu  + \frac{\eta}{\tau_\pi} \left( g^{\mu\alpha}
  u^\beta + g^{\mu\beta}  u^\alpha\right) \right] =  &
\left(1 -
  \frac{\delta_{\pi \pi}}{\tau_\pi}\right) \pi^{\alpha\beta}\, \theta
  -\frac{\delta_{\pi B}}{\tau_\pi} F^{\delta\mu} \pi^\gamma_\mu\,
\Delta^{\alpha \beta}_{\gamma\delta}\nonumber \\
  &-
  \frac{1}{\tau_\pi}\pi^{\alpha\beta}
  - \frac{1}{\omega_{\pi}}
  \left[ \left(\pi^{\alpha\lambda} u_\lambda\right) u^\beta + \left(\pi^{\beta\lambda}
  u_\lambda\right) u^\alpha \right] - \frac{1}{\omega_{\pi\pi}} \left(
  \pi^{\mu\nu} g_{\mu\nu}
  \right) g^{\alpha\beta} .
    \label{eqn:IS_shear_simple}
\end{align}
\end{widetext}

{
As a final remark we stress the ambiguity in adding or removing constraint
terms related to Eq. \eqref{eqn:constraintsPi1} and
\eqref{eqn:constraintsPi2}. While in the continuum limit Eq.
\eqref{eqn:constraintsPi2} will perfectly hold, the discrete version of
these equations might behave differently whether or not additional terms
proportional to $g^{\mu\nu}$ would be added.
One particular example would be the removal of the trace from the shear
tensor in Eq. \eqref{eqn:IS_shear_simple}. This would lead to a replacement
of the principal part of Eq.\ \eqref{eqn:IS_shear_simple} with}
{
\begin{align}
\nabla_\mu \left[  \pi^{\alpha\beta} u^\mu  + \frac{\eta}{\tau_\pi} \left( g^{\mu\alpha}
u^\beta + g^{\mu\beta}  u^\alpha  - \frac{1}{2} g^{\alpha\beta} u^\mu
\right)\right].
\end{align}
}
{
Whether or not such modifications will impact the stability of the system
will require a more in-depth analysis, which will be provided in forthcoming
work.
}

\subsection{Interpretation of the magnetic field coupling}

In the following, we would like to associate a meaning with the
  magnetic field coupling terms, $\delta_{qB}$ and $\delta_{\pi B}$.
  To facilitate this, we will draw on an analogy with the Ohm's law for a
  single fluid ideal magnetohydrodynamics in the Newtonian regime.
To this end, we consider a simple Ohm's law from Hall magnetohydrodynamics \cite{sturrock1994plasma}, 
  \begin{align}
    J^i + \omega_{g} \tau_e \varepsilon^{ijk} J_j B_k = \sigma
    E^i,
    \label{eqn:JHall}
  \end{align}
where $\omega_{g}= q/\left( m_e c B \right)$ is the electron gyrofrequency, $q$ the electron charge, $m_e$ the electron mass and $J^i$ the electric current. 
By fixing the gyration velocity $v_\perp$, we can also express this via the inverse Larmor radius $R^{-1}_L =
\omega_g / v_\perp$.\\
Taking the Newtonian limit of Eq.\ \eqref{eqn:IS_heat}, i.e. $\tau_q
  \rightarrow 0$, and going to the rest-frame of the fluid $u^\mu = \left(1, 0,0,0
  \right)$, we obtain

  \begin{align}
    q^i + \delta_{qB} \varepsilon^{ijk} q_j b_k = - \kappa \partial^i T.
    \label{eqn:qNewt}
  \end{align}
Since we only consider non-resistive plamas, we can see that the heat
  conduction current replaces the electric current compared to Eq.\
  \eqref{eqn:JHall}. In fact, if we were to consider resistive plasmas
  \cite{Denicol:2019iyh}, we would obtain exactly the same expression with
  $\sigma' e^i$ on the RHS, where $e^\mu$ is the comoving electric field,
  and $\sigma'$ the associated conductivity.

  Although Eq.\ \eqref{eqn:JHall} describes an electric and Eq.\
  \eqref{eqn:qNewt} a heat current, it can be shown that the two are
  related in a dissipative relativistic fluid description.
  Indeed, the total heat flux of the theory is given by \cite{Denicol:2019iyh}
  \begin{align}
    Q^\mu = q^\mu + \frac{ e+P}{\rho} m_b V_f^\mu,
  \end{align}
where $m_b$ is the baryon mass and $V_f^\mu$ is the particle diffusion current, which directly enters
  the electric current in Maxwell's equations \cite{Denicol:2019iyh}. 
  In defining the dissipative system in Eq.\ \eqref{eqn:Pimunu}, we have
  made a {\it frame} choice, to neglect the particle diffusion current,
  and treat the heat flux vector $q^\mu$ directly. We could have
  additionally chosen to operate in the Landau frame, and instead
  reexpressed the system in terms of the diffusion current $V_f^\mu$,
  being more similar to Eq. \eqref{eqn:JHall}.\\

  Overall, this comparison allows us to identify the coupling scale with a
  thermal gyrofrequency

  \begin{align}
    \delta_{qB} \sim \omega_{g} \tau_q.
    \label{eqn:deltaqB}
  \end{align}
  We further find it natural to identify the time scale of
  collisions $\tau_e$ with the relaxation time $\tau_q$, as those should be
  proportional to each other.
  Physically this implies that the thermal conductivity (in the
  Navier-Stokes limit) will split into a
  part parallel $\left(\kappa_{\parallel}\right)$ and perpendicular $\left(
  \kappa_{\perp}\right.$, $\left.\kappa_{\rm H} \right)$ to the magnetic field,
  where the latter term is the equivalent of the electric Hall
  conductivity \cite{Denicol:2018rbw}.
  In particular, this would lead to an anisotropic modification of the Navier-Stokes
  limit as follows{
  \begin{align}
    q^\mu \simeq& \left( \kappa_{\perp} \Xi^{\mu\nu} - \kappa_{\parallel}
    b^\mu b^\nu - \kappa_{\rm H} b^{\mu\nu}\right) 
  \left[  u^\alpha \nabla_\alpha u_\nu + \nabla_\nu \log\, T\right] T.
    \label{eqn:qNS_withB}
  \end{align}
}
  Crucially, in our current prescription these anisotropy effects are controlled by 
  a single parameter $\delta_{qB}$, which can be freely specified.
  For a more detailed discussion of this effect, see Ref. \cite{Denicol:2018rbw}.

\subsubsection{Braginskii-limit}
\label{sec:Braginskii}

{
Having discussed that the equations presented here are naturally able to
capture the effect of (Hall) anisotropies in the heat conduction, we
now turn to the strong coupling limit. It will turn out that this limit has
a very important interpretation in the limit of weakly collisional plasmas.
}

Indeed, in the strong coupling limit $\delta_{\pi B}, \delta_{q B} \rightarrow \infty$, we find that the
shear stress and heat flux must satisfy
\begin{align}
   & F^{\nu \mu} q_\mu =0,\\
   & F^{\delta\mu} \pi^\gamma_\mu\, \Delta^{\alpha \beta}_{\gamma\delta}=0.
\end{align}
Since $\pi^{\mu\nu}$ and $q^\mu$ are also subject to the constraints
\eqref{eqn:constraintsQ}-\eqref{eqn:constraintsPi2}, we find that this
imposes the following form on the dissipative fluxes and stresses,
\begin{align}
  &q^\mu = q_0 b^\mu, \label{eqn:q_brag} \\
    &\pi^{\mu\nu} = \pi_0 \left(-b^\mu b^\nu + \frac{b^2}{3}
    \Delta^{\mu\nu} \right) \label{eqn:pi_brag},
\end{align}
where $q_0$ and $\pi_0$ are Lorentz scalars, which in the Navier-Stokes limit approach
{
\begin{align}
    &q_0 \simeq - \kappa T \frac{1}{\sqrt{b^2}} \left[ b_\alpha u^\mu \nabla_\mu u^\alpha + b^\mu \nabla_\mu \log\, T\right]\,,\\
    &\pi_0 \simeq -3 \eta\, \Xi^{\mu\nu} \nabla_{\left(\mu\right.} u_{\left. \nu \right)}\,.
\end{align}
}
From a physical point of view, $\delta_B \sim R_L^{-1}\sim \omega_g \sim \lambda_{\rm mfp}$ can be identified with the
mean-free-path of the system \cite{Denicol:2018rbw} so that, as expected, in the limit of weak
collisionality $\lambda_{\rm mfp} \rightarrow \infty$ a Braginskii-like
limit is recovered \cite{braginskii1965transport} for our system of equations.

A similar formulation of relativistic viscous magnetohydrodynamics using Israel-Stewart-like equations has been
proposed for weakly collisional plasmas in Ref.\  \citep{Chandra:2015iza}. This formulation was
modelled after non-relativistic Braginskii theory \citep{braginskii1965transport}, where the shear
stress aligns with the comoving magnetic fields.
We stress that differently from the formulation of \cite{Chandra:2015iza} where the
anisotropy of the {viscous} stresses and heat flux has been 
imposed from the outset, in the framework considered here the anisotropy
emerges naturally as the gyrofrequency diverges in the limit of weak
collisionality plasmas. As such, the (truncated first-order) equations of \cite{Denicol:2018rbw}
in this limit can be considered as a generalization of Braginskii
magnetohydrodynamics
for non-resistive relativistic plasmas for \textit{finite} thermal gyrofrequencies.

\subsection{Advection terms}
\label{sec:advection}

Although we have so far managed to eliminate most gradient terms from the RHS of the equations of motion,
a set of advection and compression terms still remain.  We will now, likewise,
reformulate them as a set of implicit rate equations in the comoving frame.

We note that because of baryon number conservation, an arbitrary advected
scalar $Y$ satisfies
\begin{align}
    0= u^\mu \nabla_\mu Y = \nabla_\mu \left( \rho Y u^\mu \right).
\end{align}
If we want to enforce that $Y$ follows a certain behavior, $Y_0$, we may implicitly define an autonomous source term such that
\begin{align}
    Y\left({\bf x}, t\right) = Y_0 \left({\bf x}, t\right).
    \label{eqn:constraintY}
\end{align}
This gives rise to a relaxation current
\begin{align}
    \mathcal{I}_Y = -\frac{u^0}{\omega_Y} \left( Y-Y_0\right),
\end{align}
where $\omega_Y$ is a stiff relaxation time scale.
In order to enforce the condition \eqref{eqn:constraintY} we may add this
current to the advected part,
\begin{align}
    \nabla_\mu \left( \rho Y u^\mu \right) = \rho \mathcal{I}_Y.
\end{align}
In other words, in the local comoving frame of the fluid, this equation is
prescribing a rate equation to enforce the damping of $Y$ towards $Y_0$
on a (subgrid) time scale $\omega_Y$.
If we can evaluate the stiff current $\mathcal{I}_Y$ numerically, it can be used to replace advective derivatives of the form
\begin{align}
    u^\mu \nabla_\mu Y = \mathcal{I}_Y.
\end{align}

In the same way, we can also treat gradient terms by noting 
that
\begin{align}
    \delta^{\nu}_{\mu}\nabla_\mu Z = \nabla_\mu\left({\delta^{\mu}_{\nu} Z}\right),
\end{align}
where $\delta^{\nu}_{\mu}$ is the Kronecker symbol.
Introducing the current $Z^\nu$, we can write in non-covariant form:
\begin{align}
\partial_t\left( \sqrt{-g} Z_\nu \right) + \partial_i \left( \sqrt{-g} Z \delta^i_\nu\right) = \sqrt{-g}\Gamma^\mu_{\mu\nu} + \sqrt{-g}\mathcal{I}^Z_\nu,
\end{align}
where $\Gamma^\lambda_{\mu\nu}$ is the Christoffel symbol associated with
$g_{\mu\nu}$ and 
\begin{align}
    \mathcal{I}^Z_\nu = -\frac{1}{\omega_Z} \left( Z_\nu - Z\delta^0_{\nu}\right).
\end{align}
Here $\omega_Z$ is a stiff relaxation timescale.

Applied to our source terms this results in the following set of stiff advection equations
\begin{align}
    &\nabla_\mu \left( \rho Y_\zeta u^\mu \right) =\rho \mathcal{I}_\zeta,\\
    &\nabla_\mu \left( \rho Y_\kappa u^\mu \right) =\rho \mathcal{I}_\kappa,\\
    &\nabla_\mu \left( \rho Y_\theta u^\mu \right) =\rho \mathcal{I}_\theta,\\
    &\nabla_\mu \left( \rho Y_\eta u^\mu \right) =\rho \mathcal{I}_\eta,\\
&    \partial_t\left( \sqrt{-g} Z^\kappa_\nu \right) + \partial_i \left( \sqrt{-g} \frac{\kappa}{\tau_q} \delta^i_\nu\right) = \sqrt{-g}\Gamma^\mu_{\mu\nu} + \sqrt{-g}\mathcal{I}^\kappa_\nu.
\end{align}
The source terms of those equations are fixed according to
\begin{align}
    &\mathcal{I}_\zeta = -\frac{1}{\omega_\zeta} \left( Y_\zeta - \frac{\zeta}{\tau_\Pi} \right),\\
    &\mathcal{I}_\kappa = -\frac{1}{\omega_\kappa} \left( Y_\kappa - \frac{\kappa}{\tau_q} \right),\\
    &\mathcal{I}_\theta = -\frac{1}{\omega_\theta} \left( Y_\theta - \rho\right),\\
    &\mathcal{I}_\eta = -\frac{1}{\omega_\eta} \left( Y_\eta - \frac{\eta}{\tau_\pi} \right),\\
    &\mathcal{I}^\kappa_{\nu} = -\frac{1}{\omega_\kappa} \left( Z^\kappa_\nu- \frac{\kappa}{\tau_q}\delta^{0}_{\nu}\right),
\end{align}
where we have introduced new relaxation time scales 
$\omega_\zeta\,,\omega_\kappa\,,\omega_\eta\,,\text{and}\,\omega_\theta$.

\subsection{Summary}
In conclusion, our equations of motion for  flux-conservative dissipative magnetohydrodynamics read

\begin{widetext}
\begin{align}
  &\nabla_\mu \left( T^{\mu\nu} + \Pi^{\mu\nu} \right) = 0\,,\\
  &\nabla_\mu \left( b^\mu u^\nu - b^\nu u^\mu \right) = 0\,,\\
  &\nabla_\mu \left[ \left( \Pi + Y_\zeta \right) u^\mu
  \right] = - \frac{1}{\tau_\Pi} \Pi -  \left(1-
  \frac{\delta_{\Pi\Pi}}{{\tau_\Pi}} \right) \Pi
  \frac{1}{\rho} \mathcal{I}_\theta + \mathcal{I}_\zeta, \label{eqn:sum1}\\
  &\nabla_\mu \left[ q^\nu u^\mu + Y_\kappa T
  \Delta^{\mu\nu} \right] = -
\left( 1- \frac{\delta_{qq}}{\tau_q}\right) q^\nu\, \frac{1}{\rho} \mathcal{I}_\theta  + T \mathcal{I}_\kappa^{\nu}
 -\frac{1}{\tau_q} q^{\nu}  - \frac{1}{\omega_q} \left( q_\mu u^\mu
 \right) u^\nu-\frac{\delta_{qB}}{\tau_q} F^{\nu\mu} q_\mu,\\
  &\nabla_\mu \left[  \pi^{\alpha\beta} u^\mu  + \frac{\eta}{\tau_\pi} \left( g^{\mu\alpha}
  u^\beta + g^{\mu\beta}  u^\alpha\right) \right] = -\left(1 -
  \frac{\delta_{\pi \pi}}{\tau_\pi}\right) \pi^{\alpha\beta}\, \frac{1}{\rho} \mathcal{I}_\theta -
  \frac{1}{\tau_\pi}\pi^{\alpha\beta} 
  -\frac{\delta_{\pi B}}{\tau_\pi} F^{\delta\mu} \pi^\gamma_\mu\, \tilde{\Delta}^{\alpha \beta}_{\gamma\delta}
 \nonumber \\ &
\phantom{
  \nabla_\mu \left[  \pi^{\alpha\beta} u^\mu  + Y_\eta \left( g^{\mu\alpha}
  u^\beta + g^{\mu\beta}  u^\alpha\right) \right]=}
   - \frac{1}{\omega_{\pi}}
  \left[ \left(\pi^{\alpha\lambda} u_\lambda\right) u^\beta + \left(\pi^{\beta\lambda}
  u_\lambda\right) u^\alpha \right] - \frac{1}{\omega_{\pi\pi}} \left(
  \pi^{\mu\nu} g_{\mu\nu}
  \right) g^{\alpha\beta},\\
    &\nabla_\mu \left( \rho Y_\zeta u^\mu \right) =\rho \mathcal{I}_\zeta,\\
    &\nabla_\mu \left( \rho Y_\kappa u^\mu \right) =\rho \mathcal{I}_\kappa,\\
    &\nabla_\mu \left( \rho Y_\theta u^\mu \right) =\rho \mathcal{I}_\theta,\\
        &\nabla_\mu \left( \rho Y_\eta u^\mu \right) =\rho \mathcal{I}_\eta,\\
&    \partial_t\left( \sqrt{-g} Z^\kappa_\nu \right) + \partial_i \left(
\sqrt{-g} Y_\kappa \delta^i_\nu\right) =
\sqrt{-g}\Gamma^\mu_{\mu\nu} + \sqrt{-g}\mathcal{I}^\kappa_\nu\label{eqn:sumn}\,.
\end{align}
\end{widetext}

Before we proceed, a few comments are in order.
All second-order gradient terms $\left( \Pi \theta\,, q^\nu \theta\,,
\pi^{\mu\nu}\theta \right)$ on the RHS of all equations are treated
implicitly via the relaxation equations associated with \eqref{eqn:constraintY}. 
Although such a treatment seems approximate, compared to the
flux-divergence operator on the left-hand side (LHS), it is important to point out that
we are primarily interested in near-equilibrium behavior where
$\Pi$, $q^\nu$, and $\pi^{\mu\nu}$ are small. As such, these specific second-order
terms constitute only a minor correction and could even be omitted. This is
also consistent with neglecting all other second-order transport terms in
these equations, which should otherwise be present \cite{Denicol:2012cn}.
On the other hand, since first-order gradient terms are important for the
evolution, they require more sophisticated numerical methods, as outlined
in Sec. \ref{sec:numerics} .
Second, we point out that in deriving these equations we have made use of
our freedom to remove all terms proportional to $u^\mu$ on the RHS. Due to
the projected nature of the Israel-Stewart limit, these terms would get
removed when projecting the equations into the fluid frame (in the
continuum limit).  In the
stiff-relaxation approach, all these terms are hence implicitly accounted
for in the relaxation operator scaling with the fluid four-velocity
$u^\mu$.

It would be very interesting to investigate if our formulation of the
equations of motion can be proven to be hyperbolic in the full nonlinear
limit. This would be especially important when coupling our system to
Einstein's equations, which is needed to investigate viscous phenomena in
neutron star mergers
\cite{Shibata:2017jyf,Radice:2017zta,Alford:2017rxf,Most:2021zvc} or black
hole accretion in non-dynamical spacetimes
\cite{Foucart:2015cws,Foucart:2017axc}.  To the best of our knowledge,
there is currently no formulation of general-relativistic dissipative
magnetohydrodynamics that is proven to be strongly hyperbolic in the
nonlinear regime. For instance, such a statement is not known to hold for
the elegant approach proposed in Ref.\ \cite{Chandra:2015iza}. Drawing from
our experience with the zero magnetic field limit of Israel-Stewart-like
equations \cite{Bemfica:2019cop,Bemfica:2020xym}, such a nonlinear analysis
of hyperbolicity and causality would give rise to important constraints on
the values of the dissipative currents, the transport coefficients, and in
this case also the magnetic field. 

The equations presented in this section are written in a way that seems to
be very natural for for performing such a nonlinear investigation. Indeed,
here we have taken the first step towards this result as we have been able
to rewrite the system as a set of first-order PDEs in flux-conservative form,
which can be beneficial in such studies. We remark, however, that proving
that our nonlinear system of equations is strongly hyperbolic is a very
challenging mathematical task, which we hope to address in the near future.
On the other hand, a linearized analysis of our equations can be performed
following Ref.\ \cite{Biswas:2020rps}. The only difference would be that
the inclusion of the stiff advection equations introduces additional
non-hydrodynamic modes to the system parametrized by the new relaxation
times $\omega_\zeta$, $\omega_\kappa$, $\omega_\eta$ and $\omega_\theta$. Therefore, in
the light of \cite{Biswas:2020rps}, we leave the investigation of the
linear regime of our equations to a future dedicated study. 

\section{Numerical methods}
\label{sec:numerics}

Having derived the set of Israel-Stewart-like equations in
flux-conservative form with stiff relaxation, we now want to integrate them numerically to
investigate their behavior in a series of test problems.
One of the key features will be the treatment of the stiff source terms.
We first comment on the general numerical
approach before providing a detailed discussion of the implicit time
stepping in Sec. \ref{sec:implicit}.\\

The equations are discretized following a simple finite-volume scheme
common to many fluid dynamical problems \cite{toro2013riemann,rezzolla2013relativistic}. More specifically,
we solve a flux-conservation law of the form
\begin{align}
  \partial_t \left(\sqrt{-g}\, \mathcal{U}\right) + \partial_i \left(
  \sqrt{-g}\, \mathcal{F}^i\right) = \sqrt{-g}\, \mathcal{S},
  \label{eqn:fluxc}
\end{align}
where $\mathcal{U}$ is a conserved state vector, $\mathcal{F}^i$ is the
flux vector and $\mathcal{S}$ are the source terms.
The conserved variables, i.e. the components of $\mathcal{U}$, are given by
\begin{align}
&\rho_\ast = \rho u^0 \\
&  e_\ast = \left[ \left[e+P + b^2+ \Pi\right] \left(u^0\right)^2 +
\left(P+\Pi+\frac{b^2}{2}\right)\, g^{00}\right.\nonumber \\
& \phantom{ \tau =}\, \left. + 2 q^0 u^0 + \pi^{00} - b^0 b^0 \right)],\\
& S_i = \left[ \left(\rho h + \Pi + b^2\right)u^0 u_i + q^0 u_i + u^0 q_i
+ \pi^0_i -b^0 b_j\right]\\
& \tilde{\Pi} = \left(\Pi +
Y_\zeta\right)u^0\\
& \tilde{q}^\mu = \left(q^\mu u^0 + Y_\kappa T
\Delta^{0\nu}\right),\\
& \tilde{\pi}^{\mu\nu} = \left(\pi^{\mu\nu} u^0 +
Y_\eta \left[ g^{\mu 0 } u^\nu + g^{\nu 0} u^\mu \right]\right).
\end{align}

The precise form of the fluxes and sources depends on the spacetime
employed in the simulations and can straightforwardly be derived from 
Eq.\ \eqref{eqn:sum1}-\eqref{eqn:sumn}. {As the nature of dissipative
effects is to provide local heat fluxes and stresses, these will likely act
on scales much smaller than the curvature scale of space-time. Hence,
 for a first exploration we will conduct our numerical experiments
exclusively in flat spacetime, and leave general-relativistic test cases
for a future study. We provide an explicit
representation of the flux-conservative form of the equations in flat
Minkowski spacetime in Appendix
\ref{sec:appendix}. }\\

We discretize this set of equations by adopting a second-order accurate 
finite-volume algorithm. In particular we evolve cell averaged volume
quantities $U = \left(\Delta V\right)^{-1} \int \mathcal{U} {\rm d} V$,
where $\Delta V$ is the cell volume over which the integral is carried out.
We compute an upwinded discretization of the flux by solving the local
Riemann problem at each cell interface \cite{toro2013riemann}. In particular, we first
perform a limited interpolation step using the WENO-Z algorithm
\citep{2008JCoPh.227.3191B} for the right and left states, $U_R$ and $U_L$, at the
interface. From those, we can compute an upwinded flux adopting the Rusanov
Riemann solver \cite{toro2013riemann},
\begin{align}
  F = \frac{1}{2}\left(F_L+F_R\right) - \frac{c_c}{2} \left( U_R-U_L
  \right),
  \label{eqn:Rusanov}
\end{align}
where $c_c$ is the fastest characteristic speed at the interface.  For
simplicity, we adopt this to be the speed of light $c_c = c$.  While such a
choice leads to more diffusive numerical solutions, it is common in
relativistic simulations of non-ideal magnetohydrodynamics, when the
characteristics of the system are not known
\cite{Dionysopoulou:2012zv,Ripperda:2019lsi}.
The flux-update of the $\partial_i \mathcal{F}^i$ term is performed using a
second-order accurate discretization of the divergence operator. Especially
for (barely) resolved dissipative length scales (e.g. in the presence of
strong gradients) with smooth profiles it might be beneficial to used
improved high-order flux-update schemes
\cite{DelZanna:2007pk,mccorquodale2011high} as used in
\citep{Most:2019kfe,2018JCoPh.375.1365F}. Nonetheless, we find a
second-order scheme to be sufficient for the simple test problems
considered here.  The numerical implementation is done on top of a newly
developed version of the Athena++ framework \cite{2020ApJS..249....4S},
which utilizes the Kokkos library \citep{CarterEdwards20143202} to achieve
parallelization across modern CPU and GPU architectures.

\subsection{Implicit time-stepping}
\label{sec:implicit}
An integral part of the new formulation presented here is the enforcement
of the constraints
\eqref{eqn:constraintsQ}, \eqref{eqn:constraintsPi1} and \eqref{eqn:constraintsPi2} by means of
stiff relaxation source terms.  Due to the fact that there must be a hierarchy among the relaxation times involved, i.e., $\tau \gg \omega
\rightarrow 0$, this severe stiffness limit can only be handled using
implicit time integration schemes. One such scheme, consistent with the
explicit strong-stability preserving Runge-Kutta schemes, is the RK3-ImEx
SSP3 (4,3,3) scheme \cite{pareschi2005implicit}.  Within this scheme, we split
the source terms into two distinct contributions.  In particular we write
\begin{align}
    &\partial_t \mathcal{U}_i = \mathcal{H}_i,\\
    &\partial_t \mathcal{V}_i = \mathcal{E}_i + \mathcal{I}_i.
\end{align}
Here, we have split the hydrodynamic variables $\mathcal{U}_i =
\left(e_\ast, \rho_\ast, S_i\right)$ from the dissipative variables
$\mathcal{V}_i = \left(\tilde{\Pi}, \tilde{q}^\nu,\tilde{\pi}^{\mu\nu}\right)$.
We have further split the RHS of the dissipative variables $\mathcal{V}_i$
into explicit, $\mathcal{E}_i$, and implicit, $\mathcal{I}_i$, terms. The hydrodynamic variables remain explicit with source terms $\mathcal{H}_i$.
Explicit terms are evaluated at the current time $t^n$, while implicit
terms are evaluated at the next time step. We will give a detailed
description below.

The multistep ImEx scheme then proceeds with $n=4$ internal stages
\cite{pareschi2005implicit},
\begin{align}
    & \mathcal{U}^{\left(0\right)}_i = \mathcal{U}_i \left(t\right)\\
    &\mathcal{U}^{\left(k\right)}_i = \mathcal{U}_i + {\Delta t}\sum_{l=1}^{k-1} a_{kl} \mathcal{H}_i \left(\mathcal{U}_i^{\left(l\right) },\mathcal{V}_i^{\left(l\right)}\right)\\
    &\mathcal{V}^{\left(k\right)}_i = \mathcal{V}_i + {\Delta
    t}\sum_{l=1}^{k-1} a_{kl} \mathcal{E}_i
    \left(\mathcal{U}_i^{\left(l\right)},\mathcal{V}_i^{\left(l\right)}\right)
    \nonumber \\
&\phantom{\mathcal{V}^{\left(k\right)}_i = \mathcal{V}_i }
    + {\Delta t}\sum_{l=1}^{k} \tilde{a}_{kl} \mathcal{I}_i
    \left(\mathcal{U}_i^{\left(l\right)},\mathcal{V}_i^{\left(l\right)}\right),\\
    &\mathcal{U}_i\left(t+{\Delta t}\right) = \mathcal{U}_i + {\Delta
    t}\sum_{l=1}^{n} b_{l} \mathcal{H}_i
    \left(\mathcal{U}_i^{\left(l\right)
    },\mathcal{V}_i^{\left(l\right)}\right),
    \\
    &\mathcal{V}_i\left(t+{\Delta t}\right) = \mathcal{V}_i + {\Delta
    t}\sum_{l=1}^{n} b_{l} \mathcal{E}_i
    \left(\mathcal{U}_i^{\left(l\right)},\mathcal{V}_i^{\left(l\right)}\right)
  \nonumber \\
  &\phantom{\mathcal{V}_i\left(t+{\Delta t}\right) = \mathcal{V}_i }
  + {\Delta t}\sum_{l=1}^{n} \tilde{b}_{l} \mathcal{I}_i
  \left(\mathcal{U}_i^{\left(l\right)},\mathcal{V}_i^{\left(l\right)}\right).
\end{align}
The coefficients $a_{ij}\,, \tilde{a}_{ij}\,, b_i\,, \tilde{b}_i$ can be found in Table \ref{IMEX_butcher}.

\begin{table*}
  \begin{minipage}{2in}
\begin{tabular} {c c c c c c}
 0   & \vline & 0  &  0  &  0  & 0  \\
 0   & \vline & 0  &  0  &  0  & 0  \\
 1   & \vline & 0  &  1  &  0  & 0 \\
 1/2 & \vline & 0  & 1/4 & 1/4 & 0 \\
\hline
   &  &  0 & 1/6 & 1/6 & 2/3 \\
\end{tabular}
\end{minipage}
\begin{minipage}{2in}
\begin{tabular} {c c c c c c c}
 & $\alpha$   & \vline & $\alpha$  &  0  &  0  & 0  \\
 &0   & \vline & -$\alpha$  &  $\alpha$  &  0  & 0  \\
 & 1   & \vline & 0  &  $1-\alpha$  &  $\alpha$  & 0 \\
 & 1/2 & \vline & $\beta$  & $\eta$ & $\gamma$ & $\alpha$ \\
\hline
 &  &  &  0 & 1/6 & 1/6 & 2/3 \\
\end{tabular}
\end{minipage}

  \caption{Butcher tableau representation of the third-order IMEX-SSP3(4,3,3) scheme \citep{pareschi2005implicit}. 
  The explicit part $a_{ij}$ (left) and the implicit part $\tilde{a}_{ij}$(right) are both given in matrix form.
  The bottom row displays the coefficients $b_i$ (left) and $\tilde{b}_i$ (right). The numerical coefficients are given as follows 
   $\alpha = 0.24169426078821~,~\beta = 0.06042356519705~,~
 \eta = 0.12915286960590 ~,~ \gamma= 1/2 - \alpha -\beta -\eta$.
  }
\label{IMEX_butcher}
\end{table*}

At every substep we need to solve an implicit equation for the substep $\mathcal{V}_i^{k}$. Since the implicit matrix of Table \ref{IMEX_butcher} is lower triangular, each substep $k$, the following expression can be directly computed using the information of previous substeps,
\begin{align}
    &\mathcal{V}^{\ast\,\left(k\right)}_i = \mathcal{V}_i + {\Delta
    t}\sum_{l=1}^{k-1} a_{kl} \mathcal{E}_i
    \left(\mathcal{U}_i^{\left(l\right)},\mathcal{V}_i^{\left(l\right)}\right)
    \nonumber\\
&    \phantom{ \mathcal{V}^{\ast\,\left(k\right)}_i = \mathcal{V}_i }
    + {\Delta t}\sum_{l=1}^{k-1} \tilde{a}_{kl} \mathcal{I}_i
    \left(\mathcal{U}_i^{\left(l\right)},\mathcal{V}_i^{\left(l\right)}\right).\label{eqn:ImEx}
\end{align}

\noindent With that, the implicit equation can be written as
\begin{align}
    &\mathcal{V}^{\left(k\right)}_i = \mathcal{V}^{\ast\,\left(k\right)}_i + \alpha \Delta t \mathcal{I}_i\left(\mathcal{U}_i^{\left(k\right)},\mathcal{V}_i^{\left(k\right)}\right)\,,
\end{align}
where $a_{ii}=\alpha$. Since this equation is non-linear, numerical root-finding will be needed in order to obtain the intermediate stage $\mathcal{V}^{\left(k\right)}_i$.
\subsubsection{Solving the implicit equation}
We are now going to apply the implicit time stepping to the dissipative
magnetohydrodynamics system. 
{
  In doing so, we need to treat all source terms in Eqs.
  \eqref{eqn:sum1}-\eqref{eqn:sumn} using the ImEx method outlined
  above, see Eq. \eqref{eqn:ImEx}.
  In particular, we treat all stiff contributions proportional to
  $\tau^{-1}$ or $\omega^{-1}$ implicitly.
  We further introduce the definitions
\begin{align}
& \mathcal{I}_q^\nu = - \frac{1}{\omega_q} \left( q_\mu u^\mu
 \right) u^\nu \\
& \mathcal{I}_\pi^{\alpha\beta} =
- \frac{1}{\omega_{\pi}}
  \left[ \left(\pi^{\alpha\lambda} u_\lambda\right) u^\beta + \left(\pi^{\beta\lambda}
  u_\lambda\right) u^\alpha \right] \nonumber \\
  &\phantom{\mathcal{I}_\pi^{\alpha\beta} =} - \frac{1}{\omega_{\pi\pi}} \left(
  \pi^{\mu\nu} g_{\mu\nu}
  \right) g^{\alpha\beta}.
\end{align}
}


Since the baryon-number, energy- and momentum equations  
do not have stiff sources, the hydrodynamic variables $\rho, T,  u^\mu$
are, hence, the same in the implicit and explicit stages. 
We can then write the full set of implicit equations as
follows,
{
\begin{align}
    &\left[\Pi + Y_\zeta \right] u^0 = 
    \tilde{\Pi}^{\ast} - \alpha \Delta t
    \left[\frac{1}{\tau_\Pi} + \left( 1-
    \frac{\delta_{\Pi\Pi}}{\tau_\Pi} \right)\frac{1}{\rho}
  \mathcal{I}_\theta \right] \Pi \nonumber\\
&    \phantom{\left[\Pi + Y_\zeta \right] u^0 }
  + \alpha \Delta t \mathcal{I}_\zeta \\ 
&    q^\nu\, u^0 + Y_\kappa T \Delta^{0\nu} = \tilde{q}^{\ast\, \nu}  
    -\frac{ \alpha \Delta t}{\omega_q} \left( q^\mu u_\mu\right) u^\nu
    - \alpha \Delta t\frac{\delta_{qB}}{\tau_q} F^{\nu\mu} q_\mu\nonumber \\
&    \phantom{q^\nu\, u^0 + \frac{\kappa}{\tau_q} T \Delta^{0\nu} } 
    + \alpha \Delta t\left[ \mathcal{I}_\kappa^\nu +   \mathcal{I}_q^\nu \right],\\
&     \pi^{\alpha\beta} u^0
     = 
    \tilde{\pi}^{\ast\, \alpha \beta} - 2 Y_\eta \left( g^{0\left(\alpha\right.} u^{\left. \beta\right)} \right)  \nonumber\\ 
&   \phantom{\pi^{\alpha\beta} u^0
     =} -\alpha \Delta t\left[\frac{1}{\tau_\pi} + \left( 1-
    \frac{\delta_{\pi\pi}}{\tau_\pi} \right)\frac{1}{\rho}
  \mathcal{I}_\theta \right]\nonumber \\ &\phantom{\pi^{\alpha\beta} u^0
     =}
+ \alpha\Delta t \mathcal{I}_\pi^{\alpha\beta} 
  - \alpha \Delta t\frac{\delta_{\pi B}}{\tau_\pi} F^{\delta\mu}
  \pi^\gamma_\mu\, \Delta^{\alpha \beta}_{\gamma\delta}, \\
  & \rho u^0 Y_\zeta = \rho u^0 \tilde{Y}_\zeta^\ast -\alpha \Delta t \rho
  \mathcal{I}_\zeta, \\
  & \rho u^0 Y_\kappa = \rho u^0 \tilde{Y}_\zeta^\ast -\alpha \Delta t \rho
  \mathcal{I}_\kappa, \\
    & \rho u^0 Y_\eta = \rho u^0 \tilde{Y}_\eta^\ast -\alpha \Delta t \rho
  \mathcal{I}_\eta, \\
    & \rho u^0 Y_\theta = \rho u^0 \tilde{Y}_\theta^\ast -\alpha \Delta t \rho
  \mathcal{I}_\theta, \\
  & Z_\nu = \tilde{Z}_\nu^\ast -\alpha \Delta t \rho
  \mathcal{I}_\kappa^\nu. 
\end{align}
}
In the stiff limit of $\omega_{q/\pi/\pi\pi}\rightarrow 0$, these equations will have the following solution for a  given velocity vector $u_i$ and temperature $T$,
{
\begin{align}
&\Pi = \left( u^0 + \alpha \Delta t \Delta_\Pi  \right)^{-1}
\left[\tilde{\Pi}^\ast/u^0
- Y_\zeta^\ast\right] \label{eqn:impl_1},\\
    &q^\mu = \left( u^0 + \alpha\Delta t \Delta_q   \right)^{-1} 
    \mathcal{Q}^\mu_\nu \left[\left(q^\ast\right)^\nu - T\left(Z^\ast\right)^\nu \right], \label{eqn:impl_q}\\
    &\pi^{\alpha\beta} = \left( \left(u^0\right)^2 + \alpha\Delta t \Delta_\pi  u^0   \right)^{-1} 
    \mathcal{P}_{\gamma\delta}^{\alpha\beta}\Delta^{\gamma\delta}_{\mu\nu}
    \left(\pi^\ast\right)^{\mu\nu}, \label{eqn:impl_pi}\\
    & Y_\zeta  = \frac{\omega_\zeta}{\omega_\zeta + \alpha \Delta t} Y^\ast_\zeta
  + \frac{\alpha \Delta t}{\omega_\zeta + \alpha \Delta t} \frac{\zeta}{\tau_\Pi},\\
    & Y_\kappa  = \frac{\omega_\kappa}{\omega_\kappa + \alpha \Delta t} Y^\ast_\kappa
  + \frac{\alpha \Delta t}{\omega_\kappa + \alpha \Delta t} \frac{\kappa}{\tau_q},\\
    & Y_\eta  = \frac{\omega_\eta}{\omega_\eta + \alpha \Delta t} Y^\ast_\eta
  + \frac{\alpha \Delta t}{\omega_\eta + \alpha \Delta t} \frac{\eta}{\tau_\pi},\\
    & Y_\theta  = \frac{\omega_\theta}{\omega_\theta + \alpha \Delta t} Y^\ast_\theta
  + \frac{\alpha \Delta t}{\omega_\theta + \alpha \Delta t} \rho,\\
   & Z_\nu  = \frac{\omega_\kappa}{\omega_\kappa + \alpha \Delta t} Z_\nu^\ast
  + \frac{\alpha \Delta t}{\omega_\kappa + \alpha \Delta t} \delta^0_\nu
  \frac{\kappa}{\tau_q} \label{eqn:impl_n},
\end{align}
where we have used that
\begin{align}
    & \Delta_\Pi = \left[\frac{1}{\tau_\Pi} + \left( 1-
    \frac{\delta_{\Pi\Pi}}{\tau_\Pi} \right)\frac{1}{\rho}
  \mathcal{I}_\theta \right]\,, \\
   & \Delta_q = \left[\frac{1}{\tau_q} + \left( 1-
    \frac{\delta_{qq}}{\tau_q} \right)\frac{1}{\rho}
  \mathcal{I}_\theta \right]\,, \\
   & \Delta_\pi = \left[\frac{1}{\tau_\pi} + \left( 1-
    \frac{\delta_{\pi\pi}}{\tau_\pi} \right)\frac{1}{\rho}
  \mathcal{I}_\theta \right]\,, \\
    & \Delta t_\pi' = \alpha \Delta t \delta_{\pi B}/\tau_\pi\,, \\
    & \Delta t_q' = \alpha \Delta t \delta_{q B}/\tau_q\,, \\
    &   \mathcal{Q}^{\mu\nu} =
    \Delta^{\mu\nu}  - \frac{\left(\Delta t_q'
  \right)^2 b^2}{1+\left(\Delta t_q' \right)^2 b^2}
\Xi^{\mu\nu} 
- \frac{\Delta t_q' }{1+\left(\Delta t_q' \right)^2 b^2} F^{\mu\nu},\\
&  \mathcal{P}^{\alpha\beta}_{\gamma\delta} = b^{-2}\,
  b^\alpha b^\beta b_\gamma b_\delta + b^\alpha
  \mathfrak{p}^\beta_{\gamma\delta} + b^\beta
  \mathfrak{p}^\alpha_{\gamma\delta} +
  \mathfrak{P}^{\alpha\beta}_{\gamma\delta}
  ,\\ 
  &    \sqrt{b^2}\,\mathfrak{p}^\alpha_{\mu\nu} = 
  \left( 1 + 2 \left(\Delta t_\pi'\right)^2 b^2\right)^{-1} \left[ b_\mu
    \Xi^\alpha_{\nu} 
- \left(\Delta t_\pi'\right) b_\mu F^\alpha_{\nu} \right]\\
& \mathfrak{P}^{\alpha\beta}_{\mu\nu} = \frac{\left( 1 + 2
\left(\Delta t_\pi'\right)^2 b^2\right)}{\left( 1 + 4 \left(\Delta t_\pi'\right)^2
b^2\right)} \Xi^{\alpha}_\mu \Xi^{\beta}_\nu \nonumber \\
& \phantom{\mathfrak{P}^{\alpha\beta}_{\mu\nu}}
- \frac{ \Delta t_\pi'}{1+ 4
\left(\Delta t_\pi'\right)^2 b^2} 
\left[ F^\alpha_\mu \Xi^\beta_\nu + F^\beta_\mu \Xi^\alpha_\nu \right]
\nonumber\\
& \phantom{\mathfrak{P}^{\alpha\beta}_{\mu\nu}}
+
\frac{2 \left(\Delta t_\pi'\right)^2}{1+ 4 \left(\Delta t_\pi'\right)^2 b^2}
F^\alpha_\mu F^\beta_\nu .
\end{align}
}

\begin{figure*}[t]
    \centering
    \includegraphics[width=\textwidth]{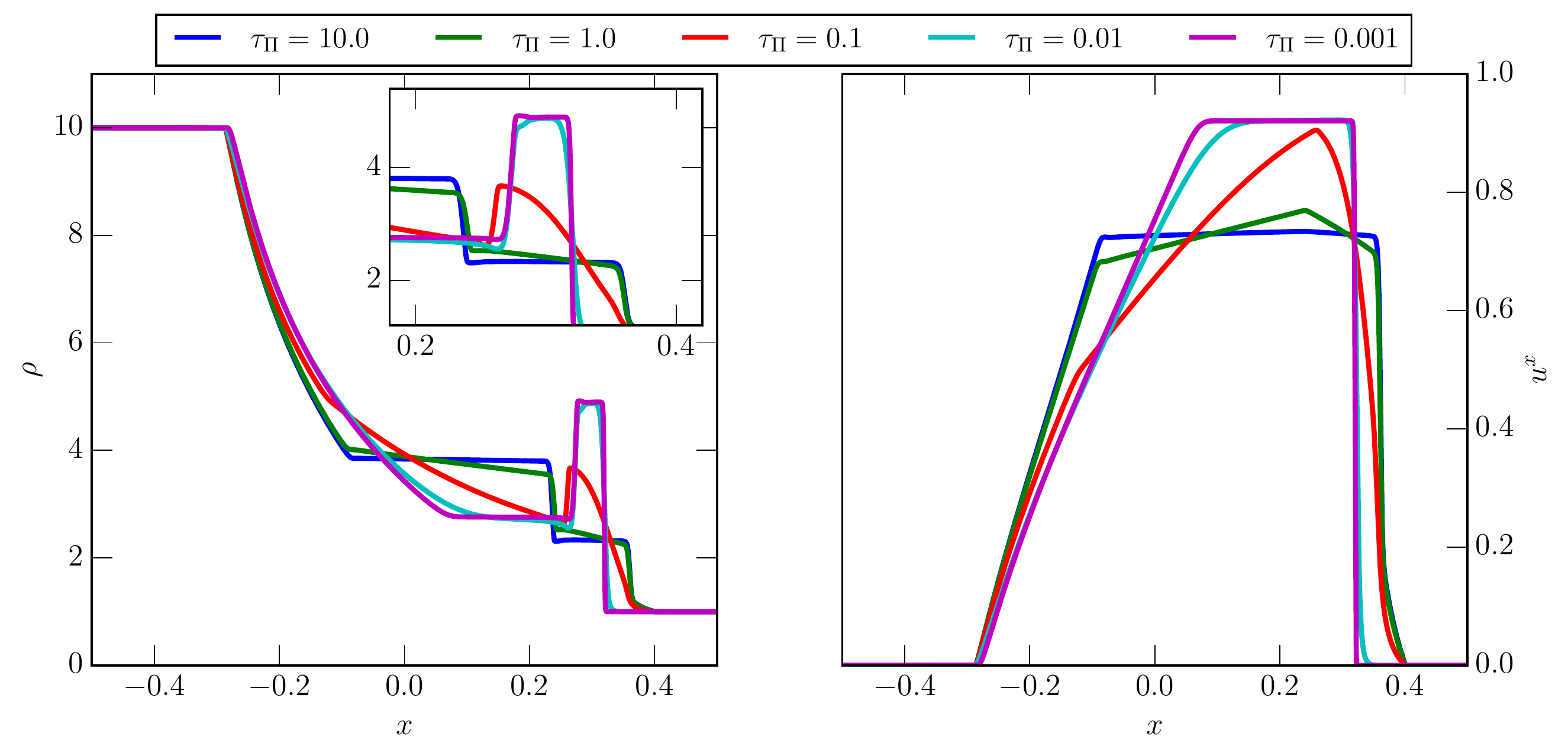}
    \caption{One-dimensional blast wave problem for different choices of
    the bulk viscous relaxation time $\tau_\Pi$ for fixed $\kappa/\tau_\Pi=1$.
    The left panel shows the rest-mass density $\rho$ at time $t=0.4$,
  the right panel the four-velocity component $u^x$.}
    \label{fig:bulk}
\end{figure*}

While those relations can straightforwardly be used when the four-velocity
$u^\mu$
and the temperature $T$ are known, this is not the case at the end of the
explicit stage, when only the conserved variables $\rho_\ast$, $e_\ast$ and $S_i$ are
given. Hence, the solution of the implicit equation needs to be embedded in
an outer root-finding loop trying to recover those variables
simultaneously. 
This problem has been studied extensively in the context of relativistic
(ideal-) (magneto-)hydrodynamics
\cite{Noble:2005gf,Galeazzi:2013mia,newman2014primitive,Siegel:2017sav,Kastaun:2020uxr},
and we refer to these studies for further details.

{
In short, our iterative procedure to solve the implicit equations proceeds as follows:
}
\begin{enumerate}
  \item 
  {
    Given the current guess for the spatial part of the four velocity
    $\bar{u}^i$ and the temperature $\bar{T}$, compute the full four-velocity $\bar{u}^0$
  from $\bar{u}^\mu \bar{u}^\nu g_{\mu\nu} = -1$ and the baryon density $\bar{\rho} =
  \rho_\ast/\bar{u}^0$.
  Together with the equation of state a full guess for the hydrodynamical
state $\left( \bar{\rho}, \bar{T}, \bar{u}^i \right)$ is then available.  \item Compute all
  transport coefficients using the above guess for the hydrodynamical
  state, compute all first-order transport coefficients $\left( \eta,
  \kappa,\zeta \right)$ and relaxation times $\left( \tau_\Pi,
  \tau_q,\tau_\pi \right)$, as well as the second-order transport terms
  $\left(\delta_{\Pi\Pi}, \delta_{qq}, \delta_{\pi\pi}\right)$ and the 
  magnetic field couplings $\left( \delta_{qB}, \delta_{\pi B} \right)$.
}
\item  {
  Using the approximate state $\left( \bar{\rho}, \bar{T}, \bar{u}^i \right)$, solve the implicit equations 
  \eqref{eqn:impl_1}-\eqref{eqn:impl_n} in order to recover the dissipative variables $\left(\bar{\Pi}, \bar{q}^\mu, \bar{\pi}^{\mu\nu}\right)$.
}
\item  {
  Next, use the approximation $\bar{\Pi}, \bar{q}^\mu, \bar{\pi}^{\mu\nu}$ to the dissipative variables obtained in the previous step to compute
  the purely hydrodynamic part of the $e_\ast$ and $S_i$,
}{
  \begin{align}
    e^{\rm hydro}_\ast =&\, e_\ast + b^0 b^0 - \pi^{00} - 2 q^0 u^0 \nonumber
    \\ &- \left(
    \frac{1}{2} b^2 + \Pi \right) \Delta^{00} - \frac{1}{2}b^2 \left( u^0
    \right)^2\,, \label{eqn:ehydro}\\
    S_i^{\rm hydro} =&\,  S_i - \pi_i^0 - q^0 u_i - q_i u^0 + b^0 b_i
    \nonumber \\ &- \left(
    \frac{1}{2}b^2 + \Pi \right) \Delta^0_i - \frac{1}{2}b^2 u^0 u_i \,,
    \label{eqn:Shydro}\\
  \end{align}
}{
  where all variables are to be interpreted as the approximate solutions obtained before, although we suppress the notation here for readability.
  These two equations can then be inverted using relations for a standard
  hydrodynamical inversion.  In particular, we use \cite{Galeazzi:2013mia},
}{
  \begin{align}
    \varepsilon &=  u^0 \frac{e_\ast}{g^{00}\rho_\ast} -
    \frac{\sqrt{S_i S^i}}{\rho_\ast} - 1\,, \label{eqn:spec_int}
  \end{align}
}{
  where $\varepsilon = e/\rho -1$ is the specific internal energy density.
}
\item{ 
  Using the equation of state, we can compute the residuals via
}{
  \begin{align}
&    {\rm res}_T =  T\left( \bar{\varepsilon}, \bar{\rho},\dots \right) - \bar{T}\,,\\
&    {\rm res}_S^i =  \frac{S_i}{\left(\bar{e}+\bar{P}\right)\bar{u}^0} - \bar{u}_i\,,
  \end{align}
}{
  where the second terms are always given by the initial guesses from Step
  1.
  Using this residual the root-finding algorithm repeats this procedure restarting at
  the first step, until these residuals are small. We typically demand that
$\max\left( \left|{\rm res}_T/T \right|, \left|{\rm res}_S^i \right| \right)<
10^{-11}$ for successful convergence.}

\end{enumerate}

{To solve the implicit equation numerically, it turned out to be
  crucial to use a stable root-finding algorithm as the Jacobian of the
  system could easily become singular. While we were able to obtain good
  results for heat conduction only with a multi-dimensional Newton-method
  \cite{broyden1965class}, 
solving the shear stresses required the use of a Newton-Krylov solver
\cite{kelley1995iterative}. 
Moreover, using a good initial guess is crucial to the successful solution
of this system. We empirically found that for initial guesses to the root-finder that
differed substantially from the true solution, the inversion converged to 
unphysical states. Differently from ideal magnetohydrodynamics inversions \cite{newman2014primitive,Siegel:2017sav,Kastaun:2020uxr}, unphysical
in this context means that the viscous degrees of freedom were unphysically
large, whereas the hydrodynamic variables were still well defined, in terms
of positive pressures and finite velocities.\\
We therefore found it beneficial to first compute an initial guess for the
Newton-Krylov solver using a standard ideal magnetohydrodynamics inversion scheme
\cite{newman2014primitive}, which neglects the viscous contributions. 
Further development and investigation will be needed to improve the
robustness and computational cost of this step, but the current approach is
sufficient to provide accurate evolutions of all test problems presented in
this work.
}

\section{Numerical tests}
\label{sec:tests}
{
Having introduced a new numerical formulation of the non-resistive viscous
relativistic hydrodynamics system, we want to  assess its ability to handle

each of the dissipative contributions. In the following, we present an
initial set of problems designed to test each of those transport
contributions individually.  Starting from one-dimensional problems for
bulk viscosity and heat conduction, we continue with a two-dimensional test
of shear viscosity and the impact of varying the thermal gyrofrequency parameter in
the presence of a magnetic field.} {For simplicity, all tests adopt a
  simplified $\Gamma$-law equation of state, i.e.  $P=\rho \varepsilon
  \left( \Gamma-1 \right)$, where $\Gamma$ will be given in the description
of each test.} As is common in numerical code testing (see e.g.
\cite{DelZanna:2007pk,2020ApJS..249....4S}) we will adopt code units
throughout.  That is, the units of all quantities are implicitly specified
relative to each other.

\subsection{Bulk viscosity}

\begin{figure*}
    \centering
    \includegraphics[width=\textwidth]{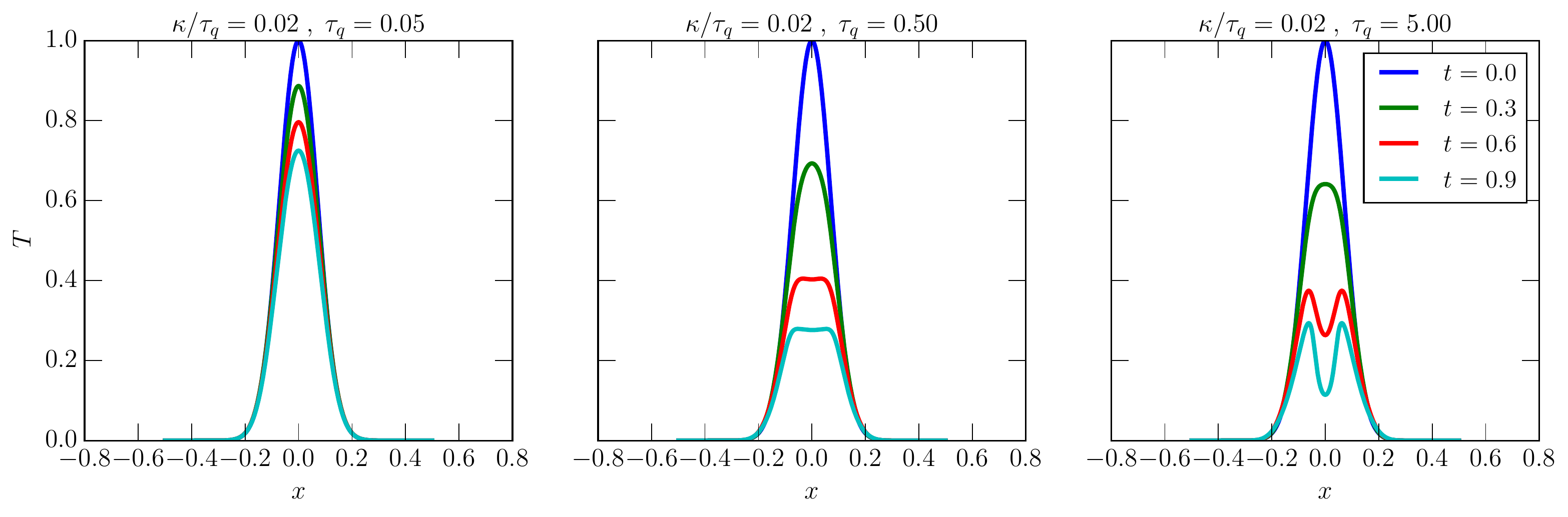}
    \caption{Thermal dissipation of an initial Gaussian temperature, $T$,
    profile. The transport coefficients $\kappa$ and $\tau_q$ are varied in
    each case. The different colors denote different times $t$ during the
  diffusion process.}
    \label{fig:heat1d}
\end{figure*}

\begin{figure}[b]
    \centering
    \includegraphics[width=0.45\textwidth]{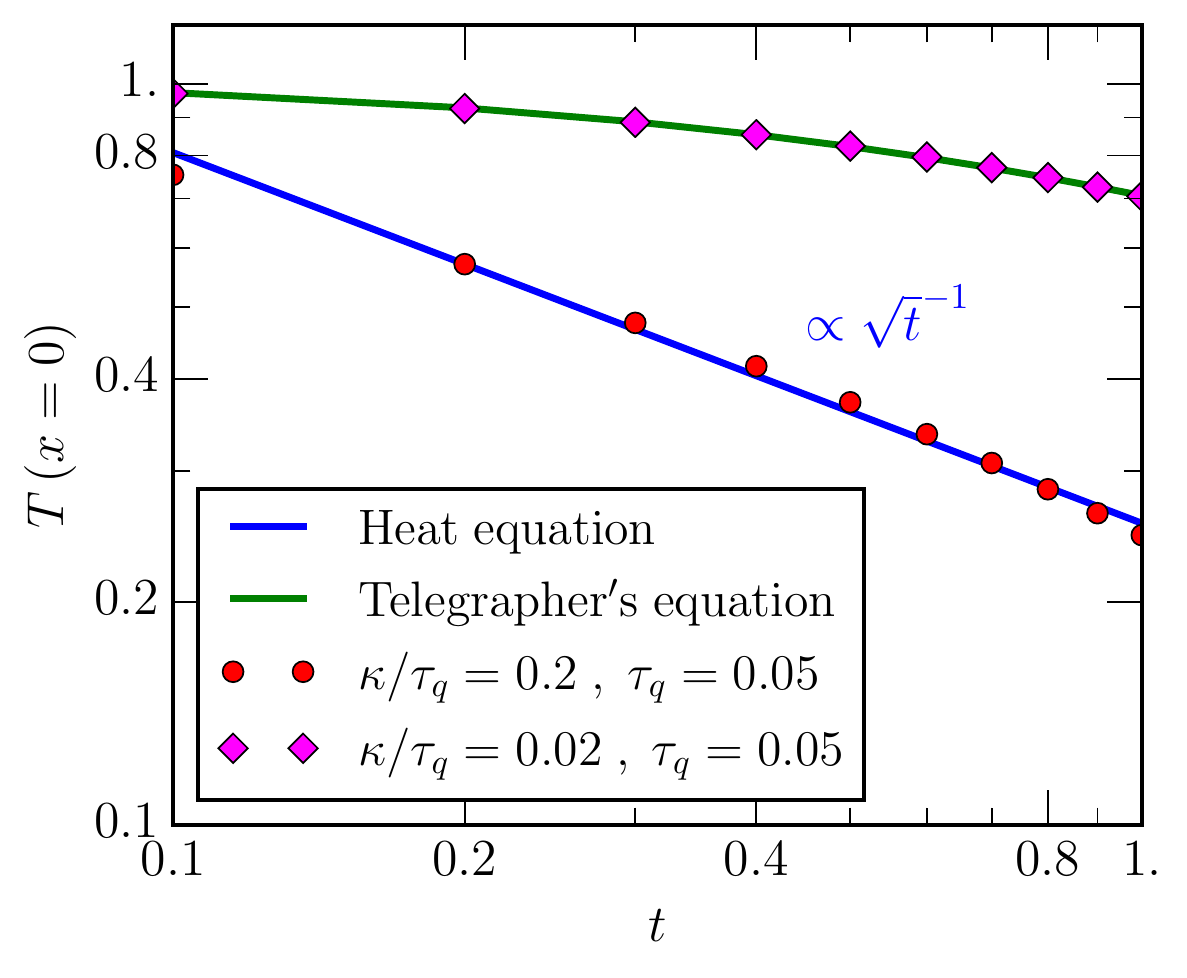}
    \caption{Comparison of the dissipation of the top, $x=0$ of a Gaussian
      temperature, $T$, profile. Shown are two evolution profiles for two sets of
      transport coefficients $\zeta$ and $\tau_q$. These are compared to
      analytic solutions of the heat equation, Eq.\ \eqref{eqn:Tgauss}, and
      the Telegrapher's equation, Eq.\ \eqref{eqn:Ttele}. }
    \label{fig:heat1d_ana}
\end{figure}

In this test we investigate the impact of bulk viscosity on a
one-dimensional relativistic shocktube problem, see also Ref. \cite{Chabanov:2021dee} for a similar test.
Following \cite{Radice:2012cu}, we adopt the following initial conditions
for a one-dimensional blast wave {launched into an ambient medium}.
{We adopt a} domain $\left[-0.5;0.5\right]$ along
the $x$-axis . All primitive variables, including the dissipative sector,
are initialized to zero. The only non-zero quantities are given separately for
$x<0$, and $x>0$ as follows :
\begin{align}
  &\rho = \left( 10^{-3}, 10^{-3} \right),\\
  & P = \left( 1, 10^{-5} \right).
\end{align}
The system is closed by adopting an equation of state with $\Gamma=5/3$.
In order to study the behavior of bulk viscous dissipation, we adopt a
fixed $\zeta/\tau_\Pi=1$  and vary $\tau_\Pi$. {This will allow us to
  probe the $\tau_\Pi\to 0$ limit discussed in Sec. \ref{sec:zero_diss} }.
  All other dissipative coefficients
are set to zero, i.e. $\eta=\kappa=0$ with their corresponding relaxation timescales
fixed to values smaller than the dynamical time of the problem, i.e.
$\tau_q=\tau_\pi =10^{-5}$ .
With this choice, the system will only be subject to bulk viscous
dissipation. 
The evolution of the baryon density $\rho$ and velocity $u^x$ is shown in
Fig.\ \ref{fig:bulk} for different values of the bulk viscous relaxation
time $\tau_\Pi$. 
We can anticipate that for this particular choice of the transport
coefficients the code will converge to two limiting solutions, see Sec. \ref{sec:zero_diss}.
In the case of $\tau_\Pi\rightarrow 0$ (magenta curve), the effective
timescale associated lengthscale $\ell_{\rm visc} \sim \zeta^{-1} = \tau_\Pi^{-1}$ of the bulk viscosity
will be too small to affect the dynamics of the shock problem, which happens
on scales $\ell_{\rm dyn} \gg 0.01$ . Hence, the solution will approach the
perfect fluid solution.
On the other hand, following the discussion in Sec. \ref{sec:zero_diss},
we approach a non-perfect fluid limit already in the case of $\tau_\Pi =10$,
which changes the shock structure (blue curve in Fig. \ref{fig:bulk}).

Intermediate values of the bulk viscosity, $\ell_{\rm dyn} \simeq \ell_{\rm
visc}$, on the other hand, have the ability to interpolate between the two
limits. This transition can be best understood when looking at the velocity
$u^x$, where for increasing relaxation time $\tau_\Pi$ the velocity
profile transitions from the higher speed perfect fluid solution (magenta
curve) to the advected solution (blue curve) at slightly lower velocities.

\subsection{One-dimensional heat conduction}
\begin{figure*}[t]
    \centering
    \includegraphics[width=0.9\textwidth]{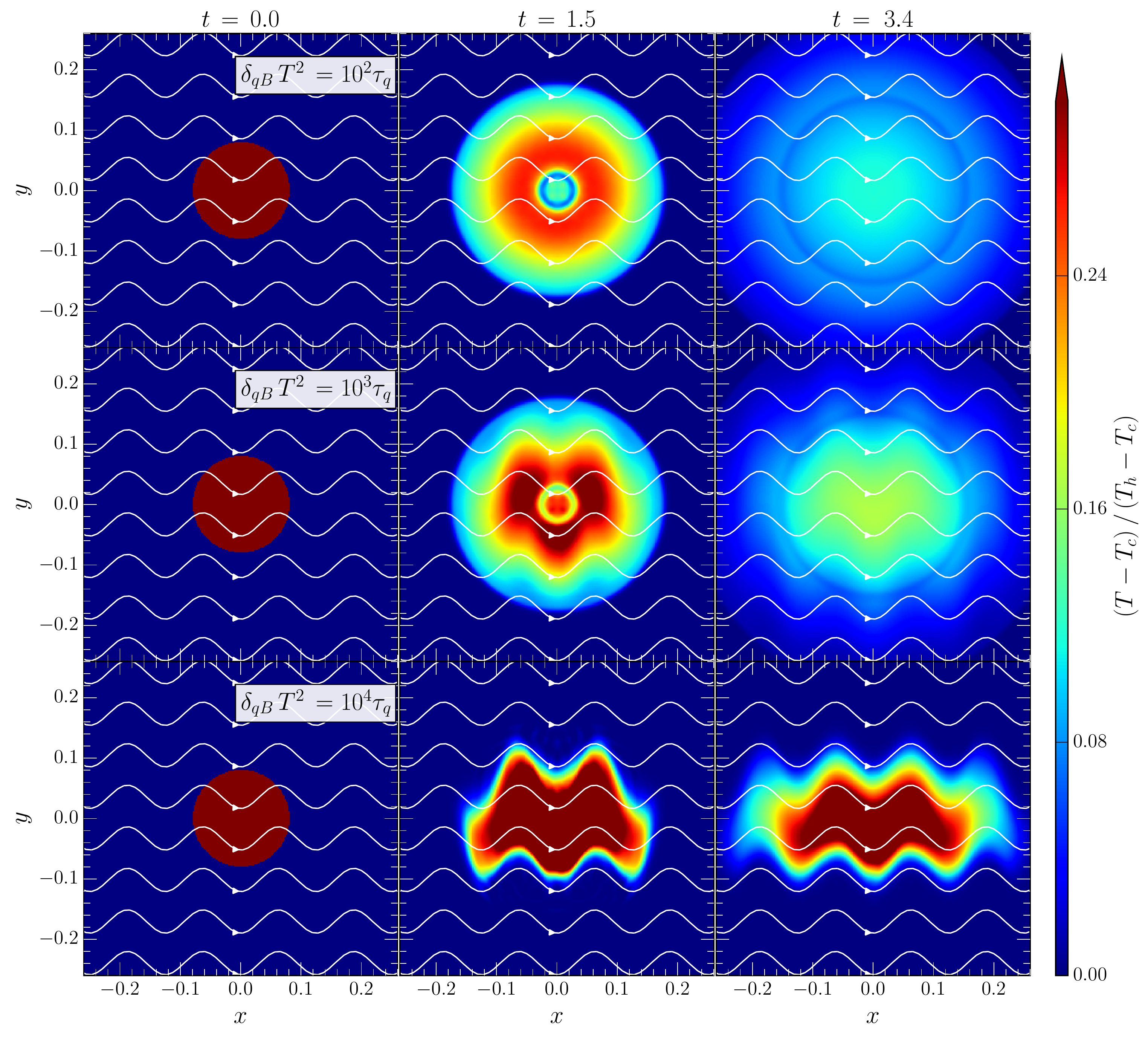}
    \caption{Anisotropic heat conductivity test. Starting from an initially
    hot inner cylinder at temperature $T_h$ surrounded by an ambient medium
    at temperature $T_c$ (left column), the evolution of relativistic
  causal heat conduction is shown for two different times in the middle and
right columns. The rows correspond to various degrees of anisotropy with
respect to the magnetic field (shown as white streamlines). We can see that
in the most anisotropic case (bottom row), heat can only flow along the
magnetic field lines. A more detailed description is given in Sec.
\ref{sec:test_heat2D}.}
    \label{fig:conductivity}
\end{figure*}

We proceed by analyzing the ability of the system to conduct heat in a
one-dimensional test setup. In order to pick a suitable set of initial
conditions, we recall that in the limit of $u^i \rightarrow 0$, the system
reduces to the Telegrapher's equation \eqref{eqn:telegraph}.
In the {Navier-Stokes} limit of $\tau_q \rightarrow 0$ we further
recover, 
\begin{align}
    \frac{\kappa m_b\left(\Gamma -1\right)}{k_B \rho} \Delta T - \partial_t T  = 0 ,
    \label{eqn:heat}
\end{align}
which is the standard heat equation. The fundamental solution to this
equation, i.e. starting from initial conditions $T\left( x,t=0 \right) = \delta\left( x
\right)$, is a Gaussian,
\begin{align}
  T\left( x,t \right) = \frac{1}{ \sqrt{4 \pi \xi t}} \exp \left( -
\frac{x^2}{4\xi t}
  \right).
  \label{eqn:Tgauss}
\end{align}
Here, we have introduced the effective diffusion constant $\xi=
\frac{\kappa m_b\left(\Gamma -1\right)}{k_B \rho}$. While this solution
only holds approximately in the limit of $\tau_q \rightarrow 0$, the full solution of
the Telegrapher's equation for the above initial condition instead reads
\cite{Romatschke:2009im},
\begin{align}
  T\left( x,t \right) =& \Theta\left( t \right)\Theta\left(
  \frac{t^2 \xi}{\tau_q} - x^2
\right)\frac{1}{\sqrt{4\xi\tau_q}}\nonumber\\
&\times e^{-
  t^2/\left( 2\tau_q \right)} I_0 \left( \sqrt{\frac{t^2}{4 \tau_q} -
  \frac{x^2}{4 \xi \tau_q}} \right),
  \label{eqn:Ttele}
\end{align}
where the Heaviside functions $\Theta$ ensure causality of the solution.
Here, $I_0$ is a modified Bessel function of the first kind.
Motivated by these observations we initialize an initial temperature
distribution to resemble a Gaussian in hydrostatic equilibrium, i.e.,
\begin{align}
  &T = \exp\left( -x^2/0.01 \right),\\
  &p = 0.1,\\
  &\rho = p/T\ \left( \text{assuming\ }k_B=m_b=1 \right),
\end{align}
where we adopt $\Gamma=4/3$.
Our simulations use a grid resolution corresponding to $N_x=1600$ grid points in the
interval $[-0.5:0.5]$. We then show the resulting profiles for different
times in Fig. \ref{fig:heat1d}. We can see that the evolution differs
drastically for fixed $\kappa/\tau_q = 0.02$ . For low $\tau_q=0.05$  and, hence,
fast relaxation (left panel), we can see that the initial Gaussian slowly
diffuses. On the other hand, for larger $\tau_q=0.5$ (center panel)  the
diffusive process happens more rapidly but the Gaussian is eventually
flattening. For even larger relaxation time, $\tau_q=5.0$ , which is much larger than the dynamical
time scale of the simulation, we see that the wave part in the
Telegrapher's equation \eqref{eqn:telegraph} takes over and the Gaussian
begins to split apart. Hence, in this limit of large mean-free path the dynamics 
is more similar to a damped wave equation than to a diffusion equation.
\\
\begin{figure*}
    \centering
    \includegraphics[width=\textwidth]{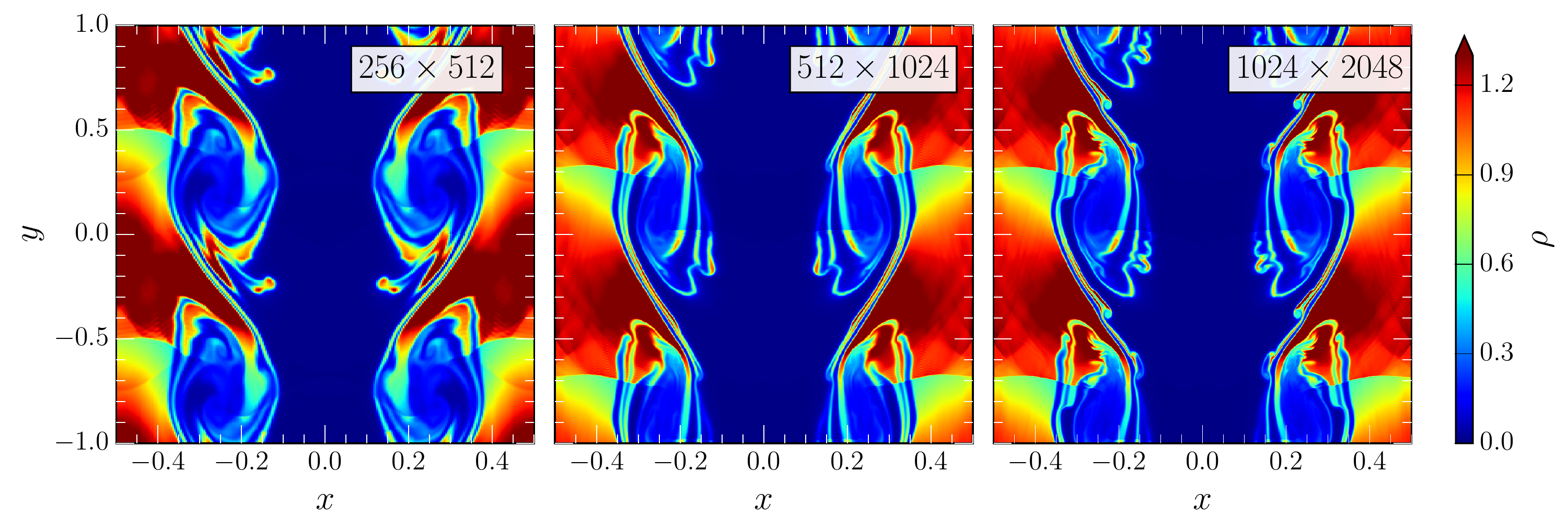}
    \caption{Two-dimensional Kelvin-Helmholtz test with heat conduction at
      $t=3.4$. Shown is the rest-mass density $\rho$. 
      In the absence of shear viscosity secondary vortices form at
    higher resolutions, denoted by the number of grid points $N_y \times N_x$.}
    \label{fig:inviscid}
\end{figure*}

In order to assess whether we are indeed recovering the correct analytic
solution of the Telegrapher's \eqref{eqn:telegraph} and Heat equations
\eqref{eqn:heat}, we compare the evolution of the center point, $T\left(
x=0, t \right)$ with the analytic solutions \eqref{eqn:Ttele} and
\eqref{eqn:Tgauss}. This comparison is shown in Fig. \ref{fig:heat1d_ana}.
We can see that for different choices of the transport coefficient
$\kappa/\tau_q$ the two limits can be reliably recovered. In particular, it
is worth highlighting that causality, which limits the instantaneous
spreading of heat, can significantly slow down dissipation compared to
standard parabolic heat conduction, for which $T\left( x=0,t \right)
\propto t^{-1/2}$.

\subsection{Two-dimensional heat conduction}
\label{sec:test_heat2D}
We now continue to explore the effects of thermal conductivity in a two-dimensional setting. 
In particular, we follow a setup first proposed by Ref. \cite{Chandra:2017auj}, 
\begin{align}
&  \rho = 
  \begin{cases}
    0.8 & \quad \sqrt{x^2+y^2} < 0.08,\\
    1.0 & \quad \sqrt{x^2+y^2} > 0.08,
  \end{cases}\\
&  P = \Gamma-1, \\
&  B^x =  10^{-4},\\
&  B^y = 10^{-4} \sin\left( 16 \pi x \right). \\
\end{align}

As before, all quantities not explicity listed have been initialized to
zero. We emphasize that the initial condition is in pressure equilibrium
and, in the absence of initial velocities, would remain static if
dissipative effects were not present.
Adopting an equation of state $\Gamma=4/3$, this corresponds to an inner
hot region with temperature $T_h = 5/12$ and an outer cold region with temperature $T_c = 4/12$.
In this problem, we explicitly include a global magnetic field to study the
dependence with the gyrofrequency parameter $\delta_{qB}$. 
Inspired by the functional form of this parameter found for ultrarelativistic
gases \cite{Denicol:2019iyh}, we chose 
\begin{align}
  &\tau_q = 0.60,\\
 &\kappa = 0.02,\\
  &\frac{\delta_{qB}}{\tau_q} T^2 = \rm const.
\end{align}
The evolution for different values of $\delta_{qB}$ is shown in Fig.\
\ref{fig:conductivity}. We can see that in the almost isotropic case, i.e.
$\delta_{qB} \rightarrow 0$, heat conduction proceeds independently of the
magnetic field geometry. More specifically, we can see in the middle and
right panels of the first row in Fig.\ \ref{fig:conductivity} that the
temperature evolution retains its cylindrical symmetry. 
With increasing degree of anisotropy (middle row), we can see that the 
temperature evolution begins to be affected by the presence of the magnetic
field. {Physically, the increase in gyrofrequency beings to suppress
cross-conductivity.} In particular, the initially cylindrical temperature profile begins
to split up (center panel) and heat conduction along the magnetic field
starts to be enhanced. Comparing the final times (right column), we still
find that the overall profile of the heat conduction is almost isotropic with only
a small degree of anisotropy being present.
Finally, by further increasing the degree of anisotropy $\delta_{qB}$, the
heat flux begins to fully align with the magnetic field, i.e. $q^\mu
\rightarrow q_0 b^\mu$, where the latter approximation is
valid in the limit of small velocities. This is the limit of extended
magnetohydrodynamics
discussed in Sec. \ref{sec:Braginskii}.

We can see from the bottom row of Fig.\ \ref{fig:conductivity} that in this
limit heat conduction across the magnetic field lines is essentially
absent and that the evolution is similar to the test case presented in Ref.\
\cite{Chandra:2017auj}. We note, however, that because the closure adopted
in the present formulation allows to dynamically adjust the degree of
anisotropy, we can naturally interpolate between the highly collisional (top row) and weakly
collisional limits (bottom row).

\subsection{Two-dimensional Kelvin-Helmholtz instability}
\label{sec:KH}

\begin{figure*}
    \centering
    \includegraphics[height=0.3\textwidth]{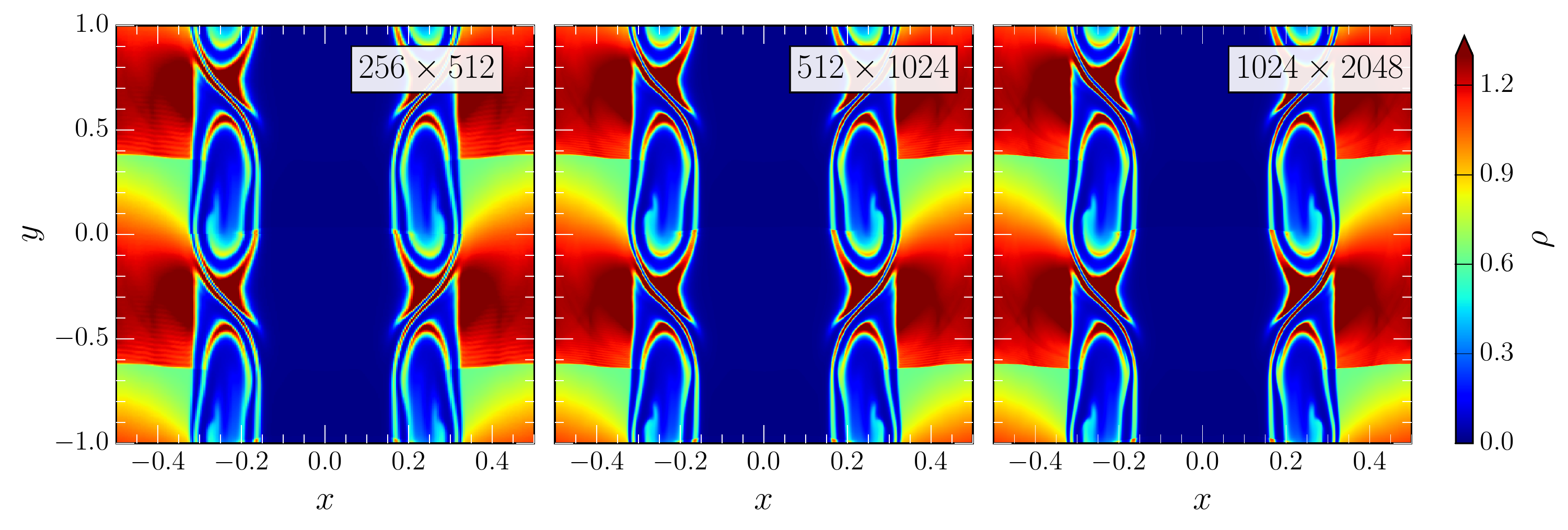}
    \includegraphics[height=0.3\textwidth]{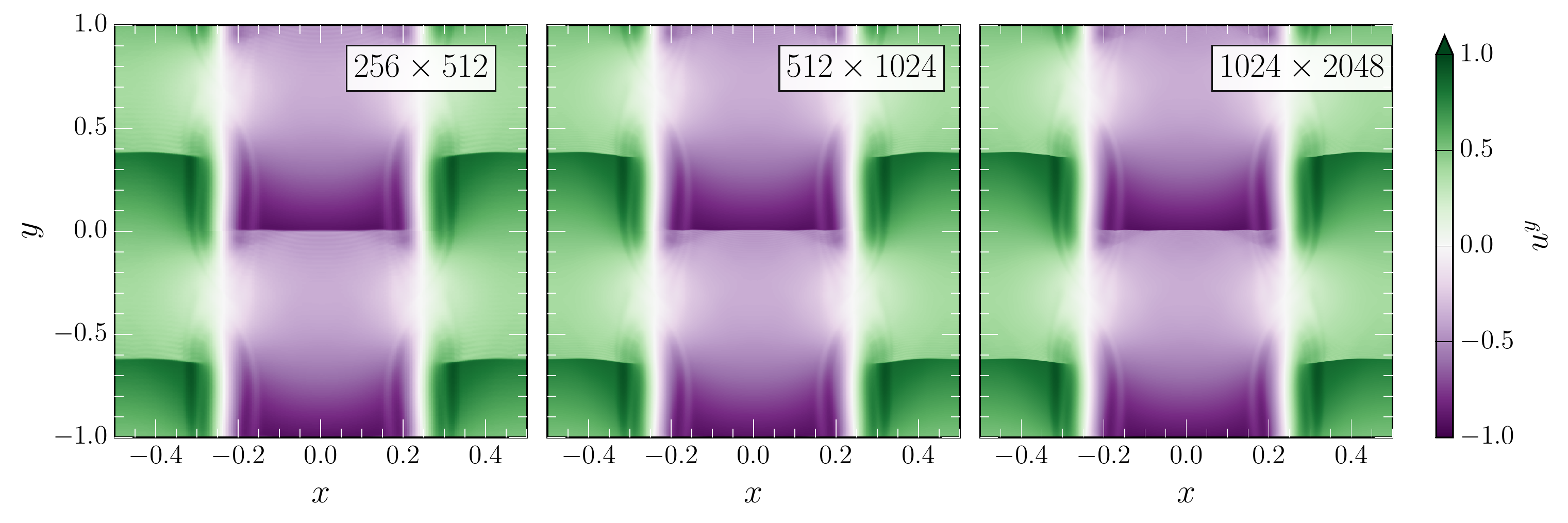}
        \includegraphics[height=0.3\textwidth]{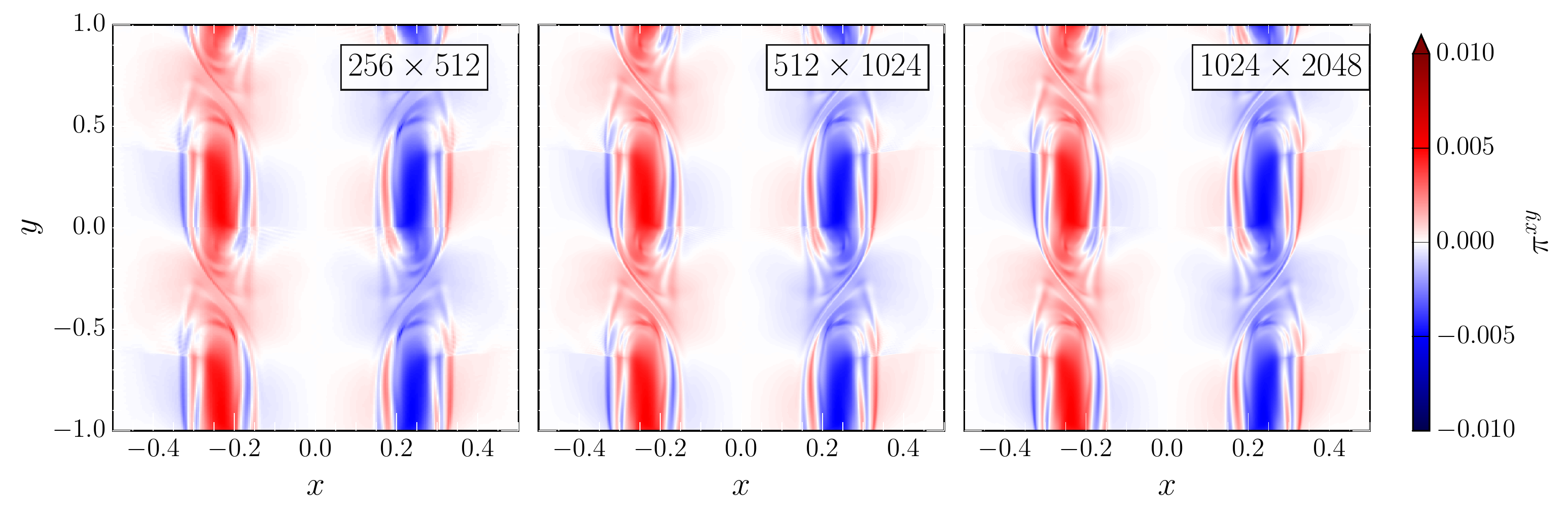}
    \includegraphics[height=0.3\textwidth]{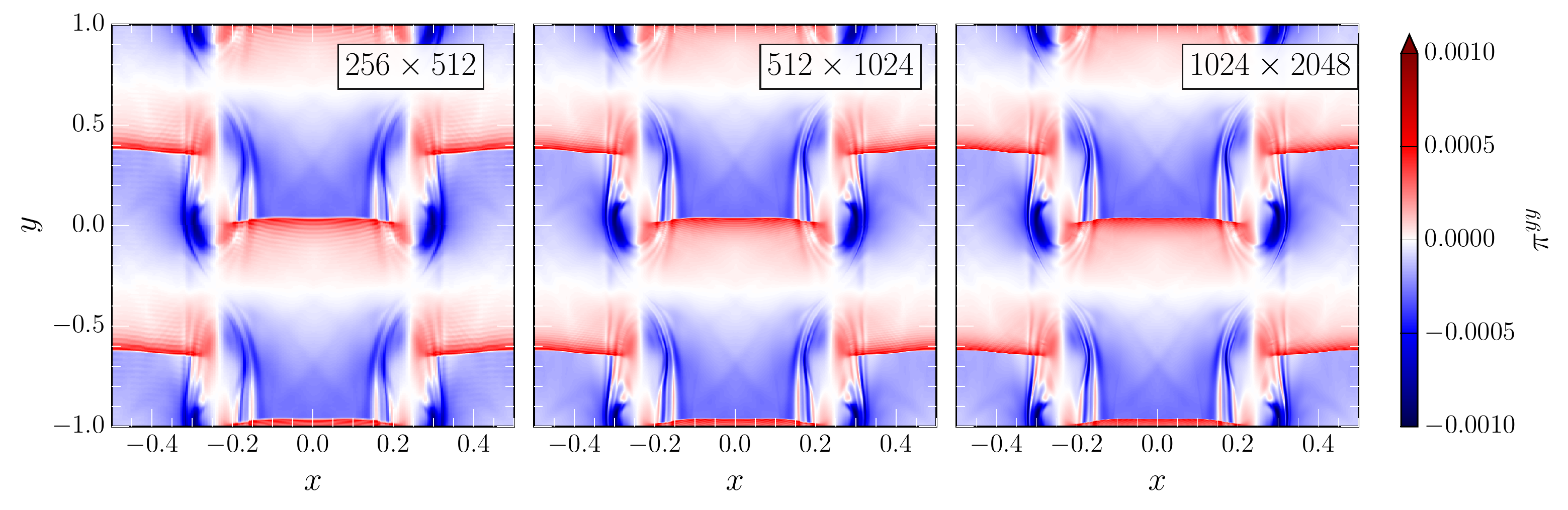}
    \caption{Two-dimensional Kelvin-Helmholtz test with heat conduction and
      shear viscosity at $t=3.4$. 
      In the presence of shear viscosity the same converged vortex structure is
      recovered also at higher resolutions, denoted by the number of grid points $N_y \times N_x$.
      Shown are the rest-mass density $\rho$ (top row), the four-velocity
      $u^y$ along the shear layer (second row), and the shear stress tensor components $\pi^{xy}$
      (third row) and $\pi^{yy}$ (bottom row).}
    \label{fig:KHhydro}
\end{figure*}

Having discussed the behavior of bulk viscosity and heat conductivity in a
series of numerical tests, we now turn to the effect of shear viscosity.
In particular, we focus on a standard test problem routinely studied in the
context of the Newtonian Navier-Stokes equations \cite{2016MNRAS.455.4274L}.
Adopting the initial conditions presented in \cite{Beckwith:2011iy}, we
study the formation of vortices in a two-dimensional Kelvin-Helmholtz
unstable shear layer. More precisely, we use 
\begin{align}
&  p= 1\,, \\
&  \rho = 
  \begin{cases}
    0.505 + 0.495\,{\rm tanh} \left[ \left( x - 0.5 \right)/0.01 \right]\,, &
    \quad x > 0\,,\\
    0.505 - 0.495\,{\rm tanh} \left[ \left( x + 0.5 \right)/0.01 \right]\,, &
    \quad x \leq 0\,,
  \end{cases} \\
  &  \frac{u^x}{u^0} = 
  \begin{cases}
    0.05\, \sin\left( 2\pi y \right) {\rm exp} \left[ -\left( x - 0.5
    \right)^2/0.1 \right]\,, &
    \quad x > 0\,,\\
    -0.05\, \sin \left( 2\pi y \right){\rm exp} \left[ -\left( x + 0.5
    \right)^2/0.1 \right]\,, &
    \quad x \leq 0\,.
  \end{cases}\\
&  \frac{u^y}{u^0} = 
  \begin{cases}
    0.5\,{\rm tanh} \left[ \left( x - 0.5 \right)/0.01 \right]\,, &
    \quad x > 0\,,\\
    0.5\,{\rm tanh} \left[ \left( x + 0.5 \right)/0.01 \right]\,, &
    \quad x \leq 0 \,,
  \end{cases}
\end{align}
where we have further adopted $\Gamma=4/3$.
As before, all quantities not listed above have been initialized to zero.
For this test problem we include both shear viscosity and heat conduction
by fixing 
\begin{align}
&  \tau_q = 0.05,\\
&  \tau_\pi = 0.1,\\
&  \kappa = 6\times 10^{-4} \tau_q,\\
&  \eta = 5\times 10^{-3} \tau_\pi, \label{eqn:etaKH}
\end{align}
which gives rise to an effective thermal Prandtl number ${\rm Pr} := \eta/\kappa
\approx 8$.\\

Shear stresses act in providing a fixed cut-off for small scale turbulence,
which is set by the length scale of the shear viscosity $\ell_{\rm shear}
\simeq \eta/\left(\rho \bar{v}\right)$, where $\bar{v}$ is a characteristic
velocity scale. If instead of an explicit viscosity this cut-off was
solely provided by the grid scale, i.e. $\ell_{\rm shear} \sim \Delta x$,
the outcome of the simulation would strongly depend on the given
resolution. Hence, demonstrating resolved and converged vortex formation 
in a Kelvin-Helmholtz unstable shear layer has become a standard test
problem for non-relativistic hydrodynamics \cite{2016MNRAS.455.4274L}.
Following the same logic, we will perform simulations at three different
resolutions, labeled by the number of grid points $N_x \times \left( 2 N_x
\right)$, where $N_x \in \left[ 256,512,1024 \right]$.

In order to establish a baseline for the effect of underresolved shear
viscosity, we perform a first set of simulations for $\eta=0$.
The resulting baryon density evolution during vortex formation is shown in Fig.
\ref{fig:inviscid}. Comparing the lowest (left panel) and highest resolutions
(right panel), we clearly find differences in the small scale evolution,
particularly inside the vortex. The presence of grid dependent effective
viscosity can best be appreciated by comparing the medium (middle panel) and high
(right panel) resolution cases. While on first glance the density
distributions look very similar, one can clearly spot the presence of
secondary vortices being formed in the shear layers of the primary vortex.
If the shear viscosity was instead constant and resolved by the numerical
resolution, we would expect convergent behavior in the vortex evolution.
That is, getting the same vortex above a certain sufficiently resolved
resolution.

To demonstrate that this is indeed the case when using our viscous scheme,
we perform the same simulation, but
with the shear viscosity $\eta$ given by Eq. \eqref{eqn:etaKH}.
The resulting evolutions for the baryon density $\rho$, the velocity
component $u^y$ along the shear layer, and the shear stresses $\pi^{xy}$ and
$\pi^{yy}$ are shown in Fig.\ \ref{fig:KHhydro}.
Comparing the density evolutions with those shown in Fig.\
\ref{fig:inviscid}, the effect of a resolved shear viscosity is strikingly
obvious. The formation of secondary vortices is suppressed at all
resolutions and the vortices have the same internal structure in all cases,
establishing that the relativistic form of the shear viscosity in
flux-conservative form is working as expected.
To better illustrate the dynamics, we next focus on the velocity $u^y$
along the shear layer (second row in Fig. \ref{fig:KHhydro}).
We can indeed see that relativistic velocities $u^y \simeq 1$ are present
in the vortices. Furthermore we can see that strong shock fronts propagate
through the vortices along the shear layer, coinciding with the 
jump (red to green color) in the rest-mass density $\rho$.
Looking at the shear stresses $\pi^{xy}$ (third row in Fig.\
\ref{fig:KHhydro}), we can see that the shear stresses are perfectly
resolved and converged at all resolutions, with the strongest stresses
present inside the vortices, as expected.

The most remarkable feature of our numerical simulations now arise in the
bottom panel of Fig.\ \ref{fig:KHhydro}. Since these simulations do not
include explicit bulk viscosity, i.e. $\zeta = 0$, the strong shock fronts
propagating along the shear layer are not resolved and should manifest
almost as discontinuities, similar to the shock tube solutions for perfect
fluids shown in Fig.\ \ref{fig:bulk}. Taking $\pi^{yy}$ as a proxy for these
stresses, we indeed find very sharp almost discontinuous jumps across the
shocks, getting sharper when going from the lowest (left panel) to the highest
(right panel) simulation. While handling such strong discontinuities might
be difficult with more traditional finite-difference approaches to the
shear stresses \cite{DelZanna:2013eua,Bazow:2016yra}, our way of incorporating the first-order
dissipative forms in flux-conservative form clearly allows us to perfectly
handle those steep jumps using high-resolution shock capturing methods.
\begin{figure*}
    \centering
    \includegraphics[height=0.4\textwidth]{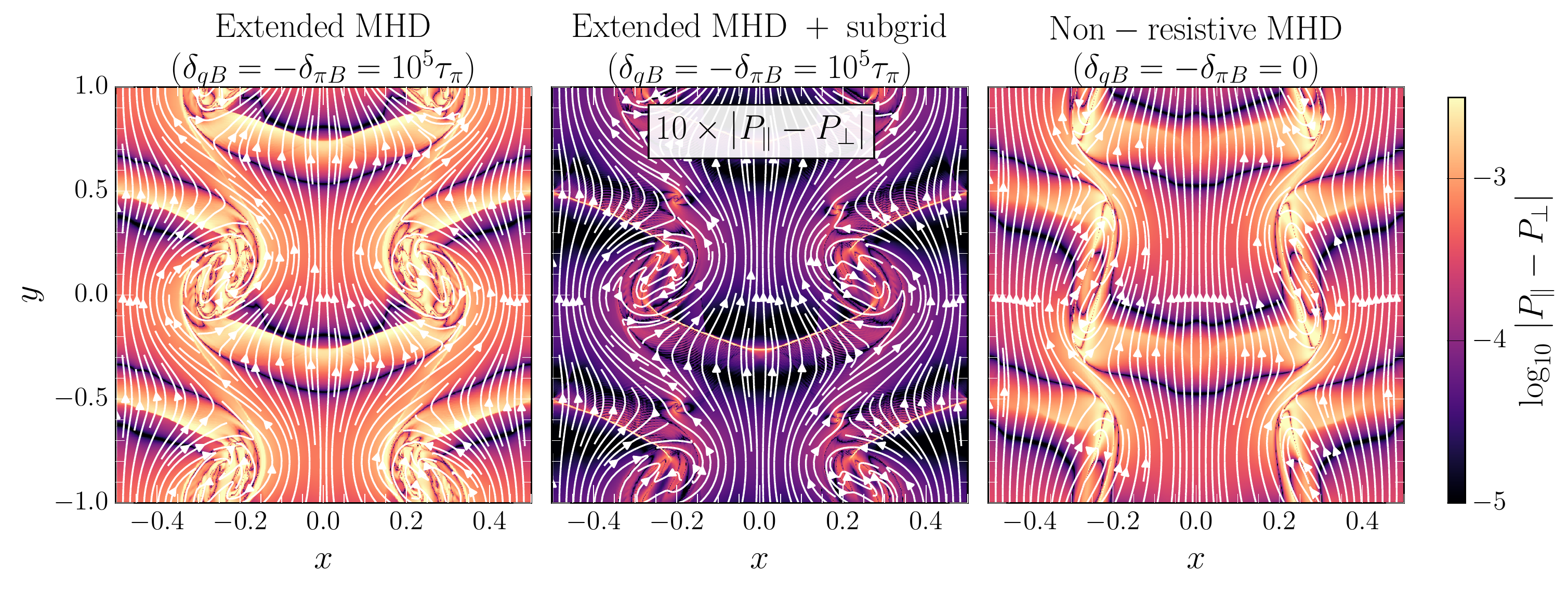}
    \includegraphics[height=0.36\textwidth]{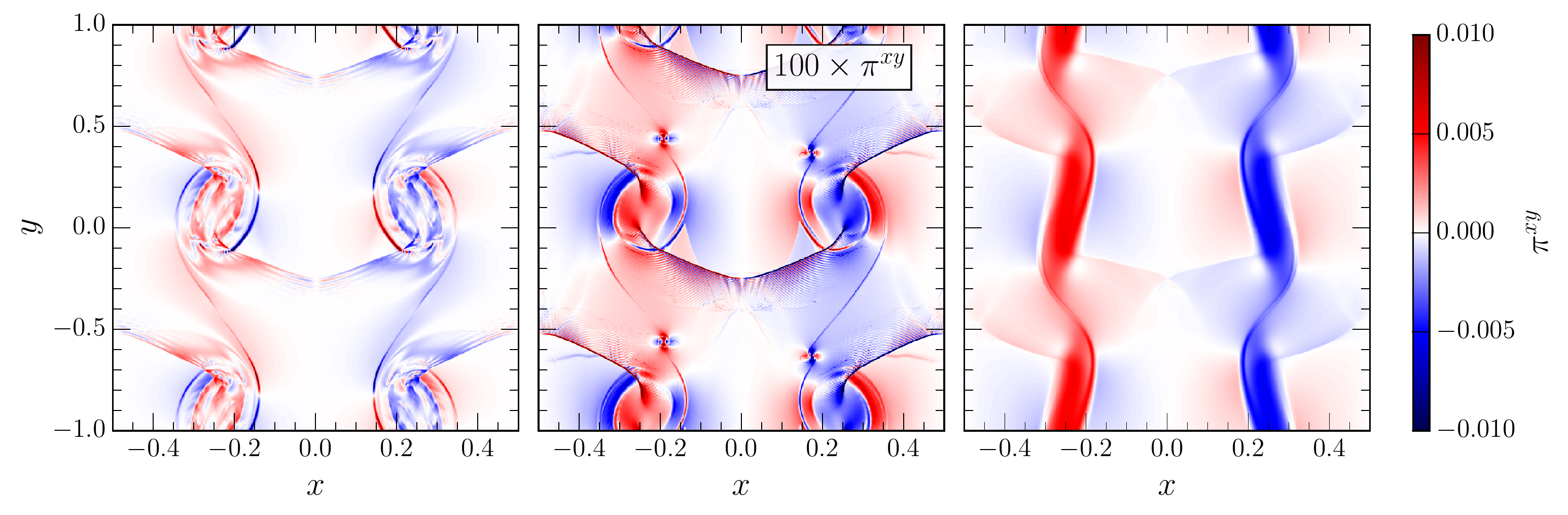}
        \includegraphics[height=0.36\textwidth]{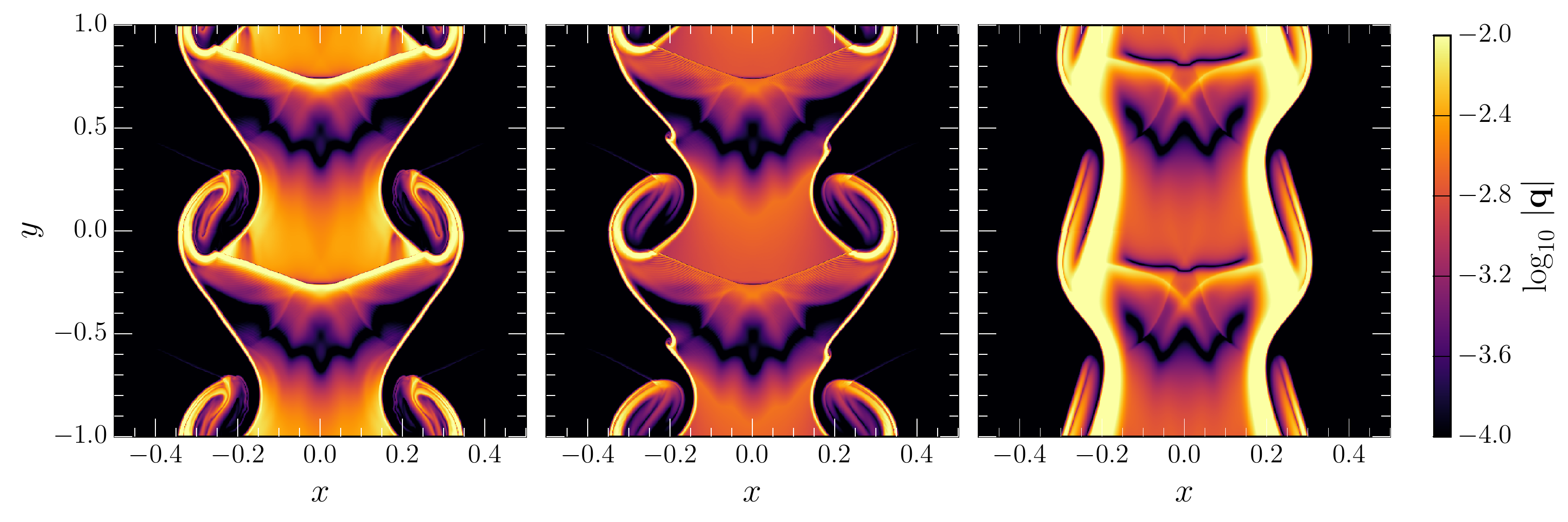}
    \caption{Two-dimensional Kelvin-Helmholtz test with anisotropic heat conduction,
      shear viscosity and finite thermal gyrofrequency at $t=3.4$.
      the left column shows results in the extended magnetohydrodynamics (Braginskii-like)
      limit of the closure. The center column in addition adds a subgrid
      model to include kinetic effects (limiting the pressure anisotropies
      according to mirror and firehose instabilities) that increase collisionality. The right column
      corresponds to a non-resistive viscous simulation.
      The rows show the pressure anisotropy $\left|P_\parallel - P_\perp \right|$ relative to the magnetic field, the in-plane shear-stresses $\pi^{xy}$ and the
    norm of the in-plane heat flux $\left|\bf q \right|$. }
    \label{fig:KHB}
\end{figure*}

\subsection{Two-dimensional Kelvin-Helmholtz instability in the presence of
magnetic fields}

As a final test, we investigate the weakly collisional regime when
shear stresses and heat fluxes are included. As discussed in Sec.\
\ref{sec:Braginskii}, in the limit of large mean-free-path the couplings 
$\delta_{qB}$ and $\delta_{\pi B}$ diverge, leading to the heat flux and
shear stresses to align with the co-moving magnetic field.\\
In the case of globally sheared accretion flows, 
hybrid-kinetic simulations of shearing flows in accretion disks
have shown that in this limit
mirror and firehose instabilities lead to a local enhancement of
collisionality \cite{2016PhRvL.117w5101K}. These instabilities are expected to kick in, once
\cite{Foucart:2015cws}
\begin{align}
&  \pi_0 > \frac{3}{2} b^2 ~ (\text{firehose instability}),\\
&  \pi_0 < \frac{3}{2} b^2 \frac{ \pi_0 + P}{\pi_0 - P } ~ (\text{mirror
  instability}). \label{eqn:mirror}
\end{align}
Additionally, the heat flux is expected to be bound
\cite{1977ApJ...211..135C,Foucart:2015cws},
\begin{align}
\left|q_0\right| \lesssim \rho c_s^3
  \label{eqn:qbound},
\end{align}
where $c_s$ is the sound speed in the fluid.
Once these instabilities are triggered they lead to a local enhancement of
collisionality and effectively pin the stresses at the threshold values for
the instability. Within the framework of dissipative hydrodynamics, this
can most easily be incorporated by locally adjusting the relaxation time
$\tau_{q/\pi} \propto \lambda_{\rm mfp}$, while keeping the ratios of the
transport coefficients fixed, i.e. $\kappa/\tau_q\,, \eta/\tau_\pi\,\simeq\, \rm const$.

To do this, we adjust the relaxation times during the implicit solve of Eq.\
\eqref{eqn:impl_q} and \eqref{eqn:impl_pi}.
We then consider the same setup presented in Sec. \ref{sec:KH}, but add a uniform magnetic field 
\begin{align}
  &  B^y = 10^{-3},
\end{align}
while keeping the other magnetic field components at zero initially.
This results in an initial plasma parameter $\beta = P/B^2 = 10^6$, 
for which the magnetic field in itself is not dynamically important.
That is, in the absence of magnetic field induced anisotropies of heat
fluxes and shear stresses, the simulation outcome will not differ from a purely
hydrodynamical simulation. However, because of the anisotropic coupling
with the magnetic field, the differences observed between the simulations
will be purely because of the anisotropic nature of the dissipative
variables.
Furthermore, for such a choice of $\beta$ the mirror and firehose
instability are expected to quench any anisotropy of the shear stresses.

We present the result of these simulations in Fig. \ref{fig:KHB} at time
$t=3.4$.  Specifically, we consider the case of $\delta_{qB}= -\delta_{\pi
B} = 10^5 \tau_\pi$, in which the equations approach the Braginskii-like
limit of extended magnetohydrodynamics \cite{Chandra:2015iza}. We also consider the same parameters, but
supplemented with the subgrid model discussed above.  Starting from the
top, we show the pressure anisotropy $\left|\pi_0\right| =
\left|P_\parallel - P_\perp \right|$, where $P_\parallel$ and $P_\perp$ are the pressures
along and perpendicular to the magnetic field. The scalar $\pi_0$ was
defined in Eq. \eqref{eqn:pi_brag}.  We can see
that the global magnetic field evolution is very similar in all cases,
except in the absence of the magnetic field coupling $(\delta_{qB}=
\delta_{\pi B}=0)$, in which the vortices are much more collapsed (right
panel). In the limit of extended magnetohydrodynamics the vortices are much more circular,
although there are differences within the vortex, when comparing the purely
extended magnetohydrodynamics case (left panel) with the subgrid modeling (middle panel).
Most importantly, we can see that the shear stresses are highly
anisotropic, especially in the case of extended magnetohydrodynamics (left). However, as
expected from the firehose and mirror instabilities, the effective subgrid
model eliminates this anisotropy almost entirely with the pressure
anisotropy being about a 100-times smaller (middle panel).

In the middle row, we show the $xy$-component of the shear stress tensor
$\pi^{xy}$. Starting from the limit of extended magnetohydrodynamics with
subgrid model (middle panel), we can see that the shear-stresses overall
are $100$ times weaker than in the other cases, as expected from the
discussion of $\left|P_\parallel - P_\perp \right|$ above.  Looking at the
case of pure Extended magnetohydrodynamics (left panel), we can also see
that small-scale substructures are present inside the vortices. Comparing
the the structures in the magnetic field (top row), we can clearly see that
shear stresses nicely align with the magnetic field geometry. When
comparing with the case without the coupling (right panel), we can see that
the stresses are uniformly present inside the vortices, where strongest
shear flows are present. As expected for an enhancement of collisionality,
the simulation including the subgrid model locally suppresses the shear
stresses inside the vortices.  The disappearance of substructure in the
main vortices is also broadly consistent with the outcome of a similar
Newtonian simulation assessing the impact of the subgrid model
\cite{Berlok:2019wki}.  Since the shear stresses are very small in this
case, they might not be able to perfectly suppress the formation of
secondary vortices, as in the isotropic viscous case, see Fig.
\ref{fig:KHhydro}. Indeed we see the (re-)appearance of secondary vortices
when adding the sugrid model (middel panel).

Finally, we turn to the heat flux present in these simulations, which is
shown in the bottom row of Fig.\ \ref{fig:KHB}.
We can see that in the case without magnetic field coupling (right
panel), a strong heat flux is present along the shear layer. 
This heat flux is able to diffuse the part of the pressure support of the
vortices helping the collapse of the vortex compared to the extended
magnetohydrodynamics
cases (left and center panels). Since for those the heat flux has to align
with the magnetic field, no heat conduction across the shear layer can
happen, as the magnetic field is aligned with the layer so that the vortices
will retain an additional pressure support.
Moreover, when considering the shock that propagates along the interface,
we can see that in the unlimited case (left panel) a strong heat flux is
present that travels with the shock. In the case of the subgrid model,
where the heat flux is clamped, see Eq. \eqref{eqn:qbound}, the transport
of heat is suppressed as these low density regions do not permit strong heat
fluxes.

\section{Conclusions}
\label{sec:conclusions}

In this work, we have presented a new formulation to numerically model
 relativitic dissipative effects in the presence of magnetic
  fields. More specifically, starting from a collisional 14-moment closure
  \cite{Denicol:2012cn} for non-resistive relativistic magnetohydrodynamics \cite{Denicol:2018rbw}, we have
  derived a set of equations suitable for the study of heat
  conduction as well as bulk and shear viscosities in an astrophysical context. 
  By treating the fluid frame projector implicitly we were able to recast the
  equations in fully flux conservative form. Different from earlier
  approaches (e.g. \cite{DelZanna:2013eua,Bazow:2016yra}), this allows us to treat all dissipative variables
  using high-resolution shock capturing methods, which removes any
  ambiguities of how to compute either explicit time or spatial derivatives
  in the dissipative sources.

In addition, the numerical scheme includes two coupling terms that
  project the shear stresses and heat flux onto the comoving magnetic
  fields. This coupling has been identified with the gyrofrequency,
  allowing us for the first time to fully include this effect in a
  relativistic magnetohydrodynamics simulations. We have further demonstrated that for large
  coupling the system approaches the relativistic Braginskii-like limit of extended
  magnetohydrodynamics investigated in Ref.\ 
\cite{Chandra:2015iza}. This allows us to study effects of weakly collisional
  plasmas that could be relevant for certain types of black hole
  accretion \cite{Foucart:2015cws,Foucart:2017axc}.
  
To investigate the properties of the solutions of the equations of motion,
we have also presented a set of numerical simulations aimed at testing each
of the dissipative effects individually. In particular, we have considered
one- and two-dimensional tests of heat conduction and viscous shear flows
in flat-spacetime. Our results showed that all
effects can be correctly captured with or without (coupled) magnetic fields
being present.  As such, we expect that this scheme will be highly suitable
to numerically study highly relativistic magnetohydrodynamical turbulence
\cite{Zrake:2011kj,Zrake:2012ek}, black hole accretion
\cite{Foucart:2015cws}, and dissipative effects in neutron star mergers
\cite{Shibata:2017jyf,Alford:2017rxf,Most:2021zvc,Hammond:2021vtv}.

  An important aspect not yet addressed in this work are causality bounds
  on the system presented here. While conditions that ensure causality in
  the nonlinear regime of Israel-Stewart-like equations formulated in the
  Landau frame (in the absence of magnetic fields or baryon density) have
  been derived in \cite{Bemfica:2020xym} (and strong hyperbolicity has been
  proven in \cite{Bemfica:2019cop} in the case of only bulk viscosity), no
  such conditions exist either when magnetic fields are included, or for
  the equations presented here, which have additional relaxation equations.
  Although our equations do approach standard Israel-Stewart-like equations
  in the stiff limit, we currently have no proof under which conditions the
  equations presented here maintain causality in the nonlinear regime (the
  linear regime of the theory derived in \cite{Denicol:2018rbw} was
  investigated in \cite{Biswas:2020rps}). Such an analysis will have to be
  performed for our system as well, although our ability to solve this
  system in a stable manner under all conditions investigated here provides
  an optimistic indication that causality and hyperbolicity can be
  established in this system \footnote{We stress at this point that a
    flux-conservative treatment requires the specification of an
    upper-limit of the system of characteristics, which we approximate with
    the speed of light. If the actual characteristics were much larger, the
  system would loose its upwind properties leading to numerical
instabilities.}.

  Beyond our current approach, several extensions are possible, in
  particular when considering applications to the hydrodynamic evolution of
  the quark-gluon plasma \cite{Romatschke:2017ejr}. In that regard, the
  next step would be to check how our framework handles Gubser flow
  \cite{Gubser:2010ze}, which has become the standard test for numerical
  codes in the field of heavy-ion collisions \cite{Marrochio:2013wla}.  In
  fact, when considering heavy-ion applications, we would have to also
  incorporate second-order coupling terms \cite{Denicol:2012cn}, though the
  effects of some of those terms on causality are not yet known in the
  nonlinear regime even at zero baryon chemical potential (for instance,
  Ref.\ \cite{Bemfica:2020xym} neglected terms involving the coupling
  between the vorticity tensor and the shear stress tensor).

Furthermore, we have limited ourselves to non-resistive plasmas. A natural
extension we are planning is the incorporation of resistivity effects,
which have been investigated in \cite{Denicol:2019iyh}. Extrapolating our
results, this will allow us to correctly treat the relativistic Hall effect
\cite{Zanotti:2011sv} and relativistic reconnection
\cite{Bransgrove:2021heo}, which could be highly relevant for flaring
process around accreting supermassive black holes
\cite{Nathanail:2020wap,Ripperda:2020bpz}.\\

\section*{Acknowledgements}
The authors thank Lev Arzamasskiy, Fabio Bemfica, Amitava Bhattarcharjee, Mani Chandra, Gabriel Denicol, Marcelo Disconzi, Charles Gammie, James Juno, Alex Pandya, Frans Pretorius, Alexander
Philippov, Bart Ripperda and James Stone for insightful discussions and
comments related to this work.
ERM gratefully acknowledges support from postdoctoral fellowships 
at the Princeton Center for Theoretical Science, the Princeton
Gravity Initiative, and the Institute for Advanced Study. 
JN is partially supported by the U.S. Department of Energy, Office of Science, Office for Nuclear Physics under Award No. DE-SC0021301.
All simulations were performed on the Helios Cluster at the Institute for Advanced Study.
\\
\bibliography{example,paper}
\appendix
\newpage
\section{Non-resistive dissipative magnetohydrodynamics in flat spacetime}
\label{sec:appendix}
The numerical tests presented in this paper are performed in flat Minkowski spacetime,
where $g_{\mu\nu} = \eta_{\mu\nu} = {\rm diag}\left(-1,1,1,1\right)$.
We can further introduce a magnetic field 
\begin{align}
  B^i =\, {^{\ast}\!F^{0i}},
  \label{eqn:Bnormal}
\end{align}
as seen by a Eulerian observer. In terms of this magnetic field, the
comoving field reads
\begin{align}
b^\mu = \left( B_j u^j, \frac{1}{u^0}\left[B^i + \left(B_j u^j\right) u^i\right]
\right).
\end{align}
The overall evolution equations of non-resistive dissipative magnetohydrodynamics in flat
Minkowski spacetime are then given by
\begin{widetext}
\begin{align}
& \partial_t\left[\rho u^0\right] + \partial_i \left[ \rho u^i\right] =0\\
& \partial_t\left[\left(\rho h + b^2 + \Pi\right) \left(u^0\right)^2  +\left(2 q^0 -
\rho \right)u^0 - \left(P+\Pi +\frac{1}{2} b^2\right) -b^0 b^0 +
\pi^{00}\right] \nonumber \\
&\phantom{\partial_t\left[\left(\rho h + b^2 + \Pi\right) \left(u^0\right)^2  \right.}
+ \partial_i \left[ \left(\rho h + \Pi +b^2\right) \left(u^0\right)^2
\frac{u^i}{u^0} +\left( q^0 - \rho\right)u^0 \frac{u^i}{u^0} +q^i u^0 + \pi^{0i}  - b^0 b^i\right] =0 \\
& \partial_t\left[\left(\rho h + \Pi + b^2\right) u^0 u_j  + q_j u^0+ q^0
u_j + \pi^0_j  - b^0 b_j\right] \nonumber \\
&\phantom{\partial_t\left[\left(\rho h + b^2 + \Pi\right) \left(u^0\right)^2\right.}
+ \partial_i \left[ \left(\rho h + \Pi + b^2 \right)u^0 u_j \frac{u^i}{u^0} + q^i u_j + q_j u^i + \pi^i_j + \delta^i_j \left(P+\Pi + \frac{1}{2} b^2 \right) - b^i b_j\right] =0 \\
&\partial_t B^j + \partial_i \left[ \frac{u^i}{u^0} B^j - \frac{u^j}{u^0} B^i\right] =0\\
&  \partial_t  \left[\left( \Pi + Y_\zeta \right) u^0
  \right]  + \partial_i  \left[\left( \Pi + Y_\zeta \right) u^0\frac{u^i}{u^0}
  \right]= - \frac{1}{\tau_\Pi} \Pi -  \left(1-
  \frac{\delta_{\Pi\Pi}}{{\tau_\Pi}} \right) \Pi
  \frac{1}{\rho} \mathcal{I}_\theta + \mathcal{I}_\zeta,\\
&\partial_t \left[ \left( q^\nu + Y_\kappa T u^\nu\right) u^0 + Y_\kappa T
  \eta^{0\nu} \right] +
  \partial_i \left[ \left( q^\nu +Y_\kappa T u^\nu\right) u^0 \frac{u^i}{u^0} + Y_\kappa T
  \eta^{i\nu} \right]= \nonumber\\
&\phantom{\partial_t \left[ \left( q^\nu +Y_\kappa T u^\nu\right) u^0 + Y_\kappa T
\eta^{0\nu} \right]}
-\left( 1- \frac{\delta_{qq}}{\tau_q}\right) q^\nu\, \frac{1}{\rho} \mathcal{I}_\theta  + T \mathcal{I}_\kappa^{\nu} -\frac{1}{\tau_q} q^{\nu}  - \frac{1}{\omega_q} \left( q_\mu u^\mu
 \right) u^\nu-\frac{\delta_{qB}}{\tau_q} F^{\nu\mu} q_\mu\\
&\partial_t \left[  \pi^{\alpha\beta} u^0  + Y_\eta \left( \eta^{0\alpha}
  u^\beta + \eta^{0\beta}  u^\alpha\right) \right] + 
  \partial_i \left[  \pi^{\alpha\beta} u^0 \frac{u^i}{u^0}  + Y_\eta \left( \eta^{i\alpha}
  u^\beta + \eta^{i\beta}  u^\alpha\right) \right]
= 
 -\frac{\delta_{\pi B}}{\tau_\pi} F^{\delta\mu} \pi^\gamma_\mu\, \tilde{\Delta}^{\alpha \beta}_{\gamma\delta}
  - \frac{1}{\tau_\pi}\pi^{\alpha\beta} 
\nonumber \\
&\phantom{
\partial_t \left[  \pi^{\alpha\beta} u^0  + \frac{\eta}{\tau_\pi} \left( \eta^{0\alpha}
  u^\beta + \eta^{0\beta}  u^\alpha\right) \right]
}
 -\left(1 -
  \frac{\delta_{\pi \pi}}{\tau_\pi}\right) \pi^{\alpha\beta}\, \frac{1}{\rho} \mathcal{I}_\theta 
 - \frac{1}{\omega_{\pi}}
  \left[ \left(\pi^{\alpha\lambda} u_\lambda\right) u^\beta + \left(\pi^{\beta\lambda}
  u_\lambda\right) u^\alpha \right] - \frac{1}{\omega_{\pi\pi}} \left(
  \pi^{\mu\nu} \eta_{\mu\nu}
  \right) \eta^{\alpha\beta}.\\
      &\partial_t \left( \rho Y_\zeta u^0 \right) +\partial_i \left( \rho
      Y_\zeta u^0 \frac{u^i}{u^0}\right) =\rho \mathcal{I}_\zeta,\\
      &\partial_t \left( \rho Y_\kappa u^0 \right) +\partial_i \left( \rho
      Y_\kappa u^0 \frac{u^i}{u^0}\right) =\rho \mathcal{I}_\kappa,\\
            &\partial_t \left( \rho Y_\eta u^0 \right) +\partial_i \left( \rho
      Y_\eta u^0 \frac{u^i}{u^0}\right) =\rho \mathcal{I}_\eta,\\
    &\partial_t \left( \rho Y_\theta u^0 \right) +\partial_i \left( \rho
    Y_\theta u^0 \frac{u^i}{u^0}\right) =\rho \mathcal{I}_\theta,\\
    &    \partial_t\left( Z^\kappa_\nu \right) + \partial_i \left( Y_\kappa \delta^i_\nu\right) =  \mathcal{I}^\kappa_\nu.\\
\end{align}
\end{widetext}

%

%
%
\label{lastpage}
\end{document}